\documentclass[journal,10pt]{IEEEtran}

\usepackage{amsfonts}
\usepackage{times}
\usepackage[pdftex]{graphicx}
\usepackage{latexsym}
\usepackage{dsfont}
\usepackage{amssymb}
\usepackage{amsmath}
\usepackage{cite}
\usepackage{verbatim}
\usepackage{subfigure}
\newtheorem{theorem}{Theorem}

\newtheorem{proposition}[theorem]{Proposition}

\def\bb0{{\mathbb{0}}}

\def\ba{{\mathbf{a}}}
\def\bb{{\mathbf{b}}}
\def\bc{{\mathbf{c}}}

\def\bg{{\mathbf{g}}}

\def\br{{\mathbf{r}}}
\def\bs{{\mathbf{s}}}

\def\bv{{\mathbf{v}}}

\def\bx{{\mathbf{x}}}
\def\by{{\mathbf{y}}}

\def\b0{{\mathbf{0}}}

\def\bA{{\mathbf{A}}}
\def\bB{{\mathbf{B}}}
\def\bC{{\mathbf{C}}}

\def\bF{{\mathbf{F}}}
\def\bG{{\mathbf{G}}}
\def\bH{{\mathbf{H}}}
\def\bI{{\mathbf{I}}}

\def\bR{{\mathbf{R}}}

\def\bU{{\mathbf{U}}}
\def\bV{{\mathbf{V}}}
\def\bW{{\mathbf{W}}}

\def\sf0{{\mathsf{0}}}

\usepackage{amsmath}
\usepackage{amsfonts}
\usepackage{amssymb}
\usepackage{graphicx}
\usepackage{subfigure}
\usepackage{setspace}

\begin{document}
\title{Cooperative Algorithms for\\ MIMO Interference Channels}
\author{Steven~W.~Peters,~\IEEEmembership{Student~Member,~IEEE,} and~Robert~W.~Heath,~Jr.,~\IEEEmembership{Senior~Member,~IEEE}
\thanks{The authors are with the Department of Electrical and Computer Engineering, 
1 University Station C0806, University of Texas at Austin, 
Austin, TX, 78712-0240 (email: \{speters,rheath\}@mail.utexas.edu, phone: (512) 471-1190, fax: (512) 471-6512).}
\thanks{This work was supported by the DARPA IT-MANET program, Grant W911NF-07-1-0028, and a gift from Huawei Technologies, Inc.}
\thanks{Copyright \copyright 2010 IEEE. Personal use of this material is permitted. However, permission to use this material for any 
other purposes must be obtained from the IEEE by sending a request to pubs-permissions@ieee.org.}}%
%         1 University Station C0806\\
%         Austin, TX, 78712-0240\\
%         Phone: (512) 471-1190\\
%         Fax: (512) 471-6512\\
%         Email: \{peters,apanah,ktruong,rheath\}@ece.utexas.edu}% <-this % stops a space

\setcounter{page}{1}
\maketitle

\begin{abstract}
Interference alignment is a transmission technique for exploiting all available degrees of freedom
in the frequency- or time-selective interference channel with an arbitrary number of users. 
Most prior work on interference alignment, however, neglects interference from other nodes in the 
network not participating in the alignment operation. This paper proposes three generalizations of interference alignment 
for the multiple-antenna interference channel with multiple users that account for colored noise, which models uncoordinated interference. 
First, a minimum interference-plus-noise leakage (INL) algorithm is presented, and shown to be equivalent to previous subspace
methods when noise is spatially white or negligible. This algorithm results in orthonormal precoders that are desirable
for practical implementation with limited feedback. A joint minimum mean squared error design is then proposed that 
jointly optimizes the transmit precoders and receive spatial filters, whereas previous designs neglect the receive spatial filter.
%This algorithm is shown to be a generalization of previous joint MMSE designs for other system configurations such as the broadcast channel. 
Finally, a maximum
signal-to-interference-plus-noise ratio (SINR) algorithm is developed that is proven to converge, unlike previous maximum SINR algorithms.
%The latter two designs are shown to have increased complexity due to non-orthogonal precoders, more required iterations, or
%more channel state knowledge than the min INL or subspace methods.
The sum throughput of these algorithms is simulated in the context of a network with uncoordinated co-channel 
interferers not participating in the alignment protocol. It is found that a network with co-channel interference can 
benefit from employing precoders designed to consider that interference, but in extreme cases, such as when only one 
receiver has a large amount of interference, ignoring the co-channel interference is advantageous. 
%Adapting between the 
%strategies in their preferred operating regimes is left to future work.
%Interference alignment is a transmission technique for exploiting all available degrees of freedom in the symmetric frequency- or time-selective interference 
%channel with an arbitrary number of users. 
%Recent work has formulated iterative algorithms for interference alignment for the case of static channel coefficients. 
%These algorithms, and interference alignment in general, are designed for the interference channel 
%in isolation with white Gaussian noise. A large network, however, is unlikely to apply interference alignment simultaneously 
%across all transceivers because of the overhead associated with channel estimation and feedback. Nodes 
%participating in interference alignment will therefore still see interference, which is often modeled as colored noise and can affect the 
%optimal precoders at finite SNR. This paper proposes three converging algorithms that account for colored noise in the MIMO interference channel
%to approximately maximize sum throughput.
%The algorithms are simulated in the context of a network with uncoordinated co-channel 
%interferers not participating in the alignment protocol. It is found that a network with co-channel interference can benefit from employing precoders 
%designed to consider that interference, but in extreme cases, such as when only one receiver has a large amount of interference, ignoring 
%the co-channel interference is advantageous. Adapting between the strategies in their preferred operating regimes is left to
%future work.
\end{abstract}
\IEEEpeerreviewmaketitle

%\begin{keywords}
%none
%\end{keywords}

% Will be ignored for other modes.
%\IEEEpeerreviewmaketitle

\newcommand{\sr}[1]{$\mathcal{#1}$}
\newcommand{\vecnorm}[2]{\left|\tilde{\bf H}_{#1}^{#2}\right|}

\section{Introduction}% (2 pages)} 
% 1. Interference channel
% 2. Interference alignment
%    - Achieves dof with infinite dimensions
%    - fixed coefficient MIMO: unknown
%    - finite dimensions practical
% 3. Prev work on this model
% 4. Our contributions (in detail)
%    - INR
%    - MMSE
%    - SINR (new)
%    - Execution of algorithms (CSI, distributed v. centralized, etc.)

%Interference is widely considered a fundamental limitation in the design of distributed wireless networks. 
%For example, distributed networks often take a time-orthogonal approach to multi-user transmission in which interference is mitigated using a 
%medium access layer (MAC) protocol such as carrier sense multiple access. For centralized networks, or networks with non-causal cooperation, 
%interference can theoretically be completely canceled, although the complexity of this strategy for a large network motivates more practical
%resource allocation to orthogonalize transmissions, usually in time, frequency, or code space.

Interference channels model a network of simultaneously communicating node pairs.  
In these channels, each transmitter has data to send to only one receiver, which also observes interference from the other transmitters in the network. 
%Interference channels are useful models for ad hoc networks and cellular networks where the wired backbone between base stations does not permit coordinated transmission.
Analysis of interference channels has shown that interference is not a fundamental limitation. 
%where distributed nodes transmit synchronously without cooperation but with channel state knowledge for every link in the network. 
In particular, with any sized interference channel with any number of users, 
the capacity for any given user will scale at half the rate of its interference-free capacity
in the high transmit power regime~\cite{CadJaf:Interference-Alignment-and-Degrees:08}.  

The key to achieving a linear capacity scaling is interference alignment (IA)~\cite{MadMotKha:Communication-Over-MIMO:08,MadMotKha:Communication-over-X-channel::06}.
With IA, interfering transmitters precode their signals to align in the unwanted users' receive space, allowing these receivers to completely
cancel more interferers than they otherwise could. The signals can be aligned in any dimension, including time, frequency, or space. 
This can be viewed as a cooperative approach because the transmitters neglect the performance of their own link to allow other users
to perfectly cancel interference.
This is in contrast to a provably suboptimal ``selfish'' approach where a transmitter 
ignores the interference it causes and aims simply to maximize its own data rate~\cite{YeBlu:Optimized-signaling-for-MIMO:03}.
Interference alignment has been shown to achieve the maximum capacity scaling, also known as degrees of freedom, of the $K$-user interference channel, 
but at finite transmit power it offers suboptimal achievable sum rate. Consequently, there is interest in finding precoders for the interference channel
that relax the perfect alignment constraint with the objective of obtaining better nonasymptotic sum rate performance.

%Methods for finding IA precoders with higher data rates at finite transmit power than in~\cite{CadJaf:Interference-Alignment-and-Degrees:08} 
Alternative IA precoder designs have been proposed for the single-antenna interference channel with time or frequency 
selectivity~\cite{SheHosVid:An-improved-interference-alignment:08,ChoJafChu:On-the-beamforming-design-for-efficient:09}. 
Closed-form IA precoders and achievable degrees of freedom for the multiple-input multiple-output (MIMO) interference channel with 
infinitely selective channels have also been found for
some asymmetric antenna arrangements~\cite{GouJaf:Degrees-of-Freedom-of-the-K-User:08}. 
Interference channels where precoding can only be done over one transmission slot are said to have constant or static coefficients. 
In this case, the degrees of freedom with linear precoding are unknown but have been hypothesized to be less than that with infinite 
selectivity~\cite{YetJafKay:Feasibility-Conditions-for-Interference:09,TreGuiRie:On-the-achievability-of-interference-alignment:09},
while non-linear precoding might achieve $KM/2$ degrees of freedom with $M$ antennas at each transmitter and 
receiver~\cite{MotGhaMad:Real-interference-alignment::09,GhaMotKha:Interference-Alignment-for-the:09}.

%A significant problem with the static MIMO interference channel is that closed-form solutions to the precoders are difficult to derive. 
A challenge in constant coefficient MIMO interference channels is that closed form solutions have been found in only a few special 
cases~\cite{CadJaf:Interference-Alignment-and-Degrees:08}. Algorithmic techniques, such as alternating 
minimization~\cite{CsiTus:Information-geometry-and-alternating:84}, have been proposed to find
precoders and explore possible
degrees of freedom for the general case~\cite{GomCadJaf:Approaching-the-Capacity-of-Wireless:08,GomCadJaf:Approaching-the-capacity-of-wireless:09,
PetHea:Interference-Alignment-Via-Alternating:09}.
Such algorithms are promising both for their ability to provide precoder solutions in a practical setting and their flexibility in application
to arbitrary networks for which closed-form solutions are unknown. The subspace algorithms 
of~\cite{GomCadJaf:Approaching-the-Capacity-of-Wireless:08,GomCadJaf:Approaching-the-capacity-of-wireless:09,
PetHea:Interference-Alignment-Via-Alternating:09}, however, still use alignment as the main objective, which is asymptotically optimal 
for the interference channel but has suboptimal throughput at finite SNR and other regimes. They also neglect colored
noise, possibly caused by co-channel interference from outside the coordinating nodes.
A maximum-SINR algorithm was proposed in~[12], but this algorithm does not optimize a global objective, assumes white Gaussian noise, and is not shown to converge.

In this paper we propose several alternative linear precoding designs for MIMO interference channels. While maximizing the sum rate is the primary
objective, we do not directly maximize sum rate due to analytical intractability. Instead we approximate a sum rate maximization via algorithms
with varying performance and complexity tradeoffs.
%aims to find sum rate maximizing precoder solutions for MIMO interference channels with static coefficients. 
%Finding precoders that directly maximize sum rate for any given channel realization is intractable, however, so the paper
%proposes metrics to approximate a sum rate maximization. 
First, we derive a generalization of subspace alignment that includes colored noise, which biases the preferred alignment subspaces.
The resulting objective, which minimizes the interference plus noise leakage (INL), 
results in orthogonal precoders amenable to quantized CSI. This algorithm is shown to be a special type of minimum mean squared error (MMSE) design, 
and at high SNR or white noise at all receivers is shown to reduce to the IA subspace
methods of~\cite{GomCadJaf:Approaching-the-Capacity-of-Wireless:08,PetHea:Interference-Alignment-Via-Alternating:09}.
From this, interference alignment is shown to be an MMSE-type solution at infinite SNR, where interference-suppression filters are optimal.
As with previous forms of interference alignment, the proposed minimum INL algorithm does not consider
the signal power at any given user and is thus suboptimal with finite transmit power. Further, this algorithm and previous designs derive precoders but neglect
receiver design, which could be optimized jointly with the precoders.

Inspired by the connection between mutual information and mean-square error~\cite{GuoShaVer:Mutual-information-and-minimum:05}, 
we derive an explicit joint MMSE precoder/receiver design for the interference channel. 
Although it does not directly maximize the sum rate, the joint-MMSE design results in a higher sum rate than subspace methods.
It does not lead to orthonormal precoders, making quantized feedback design more difficult. 
The MMSE design is shown to be a generalization of previous approaches for point-to-point and multiuser 
settings~\cite{Sal:Digital-transmission-over:85,YanRoy:On-joint-transmitter-and-receiver:94,SamPau:Joint-transmit-and-receive:99,
TenAdv:Joint-multiuser-transmit-receive:04,ZhaWuZho:Joint-linear-transmitter:05}. 
Further, the design is more computationally complex and requires more iterations at high SNR than subspace designs. MMSE-based designs have also recently been developed independently
in~\cite{SchShiBer:Minimum-mean-squared:09,SheLiTao:The-new-interference-alignment-scheme:10}.

To more directly optimize the sum rate we formulate a maximum SINR algorithm, which is proven to converge via alternating minimization of a global performance function.
The maximum SINR algorithm derived in~\cite{GomCadJaf:Approaching-the-Capacity-of-Wireless:08} is 
shown to be an approximation to the one derived in this paper. On average, the two are shown to have the same performance, but for any given channel 
realization may result in different sum rates.
This design often has increased throughput relative to MMSE and subspace approaches, but finds nonorthogonal precoders and requires more
channel state information if run in a distributed manner. 

In summary, this paper proposes three algorithms that span the tradeoff between performance and complexity for the static MIMO interference channel.
The minimum INL algorithm has the same complexity as previous work but has improved performance when colored noise exists at any
receiver. The joint-MMSE design has further rate enhancements regardless of the noise covariance matrices but has a computationally more complex optimization
and non-orthogonal precoders. The maximum SINR design has the best overall performance of all proposed strategies (in most cases, as shown in the simulations),
but requires more channel state information than the previous designs and also results in non-orthogonal precoders which are difficult to quantize in a practical
setting~\cite{LovHeaStr:Grassmannian-beamforming-for-multiple-input:03}. The proposed algorithms are then simulated alongside existing methods in regimes 
previously unconsidered in the literature. 
For example, the algorithms are simulated in an environment with an uncoordinated interferer that is not participating in the alignment protocol. 
This colors the noise at each receiver, and if its power is scaled with the rest of the transmitters, resulting in reduced capacity scaling. Each of the algorithms can 
outperform the others in different regimes, and each of these regimes is simulated and enumerated.

The rest of this paper is organized as follows: Section~\ref{sec:model} presents the system model under consideration; 
%Section~\ref{sec:int_align} gives background on %interference alignment for the interference channel; 
Section~\ref{sec:algorithms} presents the new MMSE and INL algorithms and derives a
maximum SINR algorithm with proven convergence and analyzes each of the methods; 
Section~\ref{sec:sims} presents simulations under uncoordinated interference and colored noise; and Section~\ref{sec:conclusion} 
concludes the paper and gives directions for future work.

Before proceeding, we introduce notation. The $\log$ refers to $\log_2$. Bold uppercase letters, such as ${\bf A}$, denote 
matrices, bold lowercase letters, such as ${\bf a}$, denote column vectors, and normal letters $a$ denote scalars.
The letter $\mathbb{E}$ denotes expectation, $\mathbb{C}$ is the complex field, $\mathbb{R}\{a\}$ is the real component of 
complex scalar $a$, $\min\{a,b\}$ denotes the minimum of $a$ and $b$, 
$\nu_{\rm min}^R\left(\bA\right)$ is the matrix whose columns are the eigenvectors corresponding to the $R$ smallest eigenvalues of matrix 
$\bA$, $\mathrm{tr}\left(\bA\right)$ is the trace of matrix $\bA$, $|a|$ is the magnitude of the complex number $a$, 
$\|\bf a\|$ is the Euclidean norm of vector ${\bf a}$, and $\left|\bA\right|$ is the
determinant of square matrix $\bA$. $\bA^*$ is the Hermitian transpose of matrix $\bA$ and $\bA^{-1}$ is its inverse. 
The matrices $\bI$ and $\bf 0$ are the identity matrix and all zero matrix, respectively,
of appropriate dimension. Finally, we use $\{\bF_\ell\}$ when referring to the set of precoders and $\bF_\ell$ when referring to the precoder
at transmitter $\ell$, and similarly for receive spatial filters $\{\bG_k\}$.

%%%%%%%%%%%%%%%%%%%%%%%%%%%%%%%%%%
\section{System Model}\label{sec:model}
\begin{figure}
\centering
\includegraphics[width=3.5in]{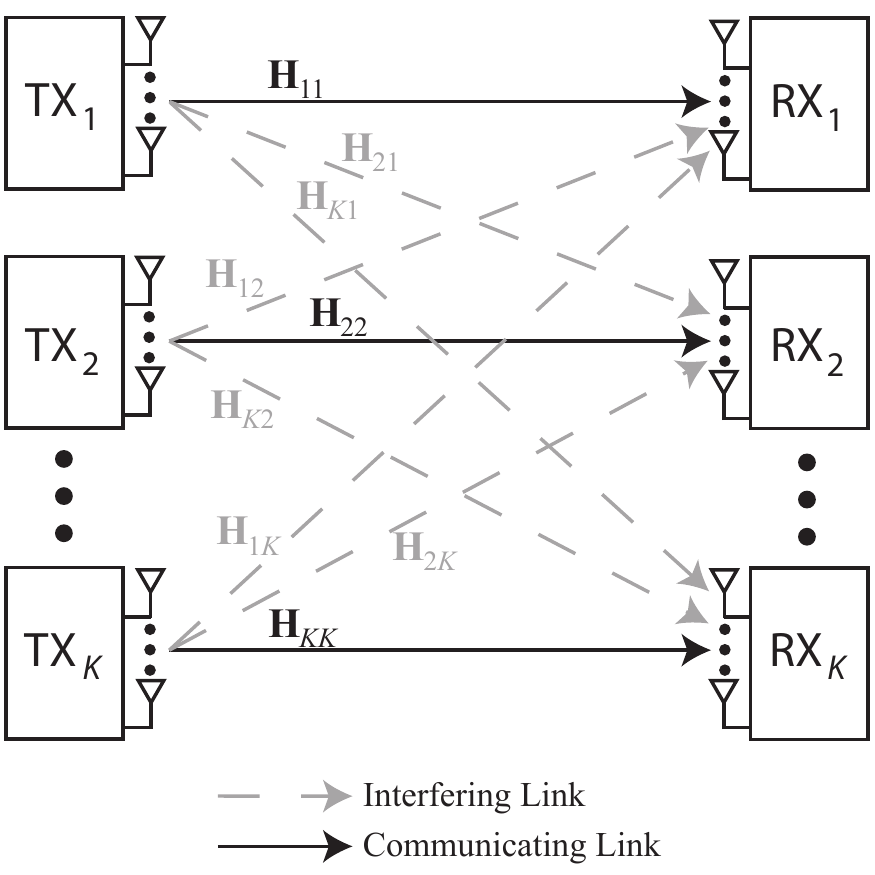}
\caption{The MIMO interference channel. Each transmitter is paired with a single receiver, and all links are non-negligible.}
\label{fig:channel_model}
\end{figure}

Consider the $K$-user MIMO interference channel illustrated in Figure~\ref{fig:channel_model},
with $K$ transmit-receive pairs. A wireless channel links each receiver to each transmitter, but a given transmitter
only intends to have its signal decoded by a single receiver. The $k$th transmitter possesses $M_k$ antennas with which to transmit 
$S_k\le M_k$ spatial streams, and the $k$th receiver (which
is to decode the signal from the $k$th transmitter) possesses $N_k\ge S_k$ antennas. In some analysis and simulations, all
users will have the same antenna configurations so that
$M_k=M, N_k=N$, and $S_k = S, \forall k$; we denote this symmetric case as an $(M,N,K)$ interference channel with $S$ streams per user. 

This paper considers the narrowband MIMO interference channel where each link is static for the duration of a transmission, 
but may change between successive transmissions. This is the block fading model where all the links in the network
are constant for the period of transmission, creating a tractable approximation to more realistic continuous fading models. 
Linear precoding is done independently over each channel realization, favoring simplicity over the possible degrees of freedom gained by jointly precoding
over realizations.
%Rather than jointly precoding over the different realizations, 
%which can achieve asymptotically the degrees of freedom of the interference channel while increasing the 
%delay or bandwidth~\cite{CadJaf:Interference-Alignment-and-Degrees:08} and possibly requiring non-causal knowledge of the channel state, our
%model focuses on lower transmission complexity, causal channel state knowledge, and realistic delays. 
This is the same model as previous work on algorithms for the interference 
channel~\cite{GomCadJaf:Approaching-the-Capacity-of-Wireless:08,GomCadJaf:Approaching-the-capacity-of-wireless:09,
PetHea:Interference-Alignment-Via-Alternating:09}. %in which the channel 
The transmission of all $K$ users is synchronized such that each begins and ends each transmission simultaneously, and no frequency
offsets exist in the network. 
We therefore take the standard approach~\cite{GomCadJaf:Approaching-the-Capacity-of-Wireless:08,PetHea:Interference-Alignment-Via-Alternating:09} 
and focus on the transmission of a single vector symbol $\bs_k$ from transmitter
$k\in\{1,\dots,K\}$, neglecting any time dependency. 

Transmitter $k$ uses linear precoder $\bF_k\in\mathbb{C}^{M_k\times S_k}$ to map $S_k$ symbols in $\bs_k$ to its $M_k$ transmit antennas,
\begin{equation}
\bx_k=\bF_k\bs_k,
\end{equation}
where 
the transmitted symbols are i.i.d.~such that $\mathbb{E}\bs_k\bs_k^*=\bI$, 
the precoder is normalized such that $\|\bF_k\|^2_F\le \rho_k$,
and $\rho_k$ is the transmit power at transmitter $k$.
Receiver $k$ observes the signal 
\begin{equation}
\by_k=\sum_{\ell=1}^K\bH_{k,\ell}\bF_\ell\bs_\ell + \bv_k,
\label{eq:y_k}
\end{equation}
where 
%$\alpha_{k,\ell}$ is the path loss and fading coefficient between transmitter $\ell$ and receiver $k$, 
$\bH_{k,\ell}$ is the channel between transmitter $\ell$ and receiver $k$ 
%with $\|\bH_{k,\ell}\|_F = 1$, 
and $\bv_k$ is Gaussian noise at receiver $k$ with spatial covariance matrix $\bR_k=\mathbb{E}\bv_k\bv_k^*$.
For the analysis in this paper we assume that the channels $\{\bH_{k,\ell}\}$ are each full rank and mutually independent, 
the transmitters send independent data ($\mathbb{E}\bs_k\bs_\ell^*={\bf 0}$ for $k\ne\ell$) 
and all transmitted signals are statistically independent from the noise at any receiver ($\mathbb{E}\bs_\ell\bv_k^*={\bf 0}$ for all $(k,\ell)\in\{1,\dots,K\}^2$).
%For notational convenience, define $\gamma_{k,\ell}=\alpha_{k,\ell}\rho_k$. 
No assumptions are made on the noise power or covariance at any receiver. 
Rewriting~(\ref{eq:y_k}), receiver $k$ sees
\begin{equation}
\by_k=\bH_{k,k}\bF_k\bs_k + \sum_{\substack{\ell=1\\\ell\ne k}}^K\bH_{k,\ell}\bF_\ell\bs_\ell + \bv_k.
\label{eq:y_k2}
\end{equation}
The vector $\bs_k$ is the signal to be decoded by receiver $k$, and the summation term in~(\ref{eq:y_k2}) is called \emph{coordinated interference}, since it is
caused by transmitters that may coordinate to minimize its effect. 
%Such decoding
%could be done in various ways. The most straightforward way of decoding $\bs_k$ would be a maximum
%likelihood (ML) approach, which would consist of an exhaustive search over the set of possible transmitted vectors from each
%transmitter. This is a prohibitively complex solution for all but the smallest networks and constellations~\cite{Ver:Multiuser-Detection:98}. %; in a $K$-user system where 
%%each transmitter sends $S$ streams, each with $b$ bits per symbol, ML decoding would require a search over $2^{bSK}$ symbols.
%Receiver $k$ could use any multiuser detector to decode or remove interference, but these will not increase the 
%degrees of freedom over a linear receiver~\cite{CadJaf:Interference-Alignment-and-Degrees:08}. 
%While nonlinear receivers may have advantages in regimes where interference
%is much stronger than the desired signal, they are beyond the scope of this paper. Thus, we focus on linear receivers for analytical
%tractability. 
%%Another possible method for decoding $\bs_k$ is successive interference cancellation (SIC). 
%
%We consider the case where receiver $k$ applies a linear spatial equalizer $\bG_k$ to its received signal $\by_k$ for decoding $\bs_k$. 
%%Mathematically,
%The signal after spatial equalization is denoted
%\begin{equation}
%\bz_k = \bG_k^*\by_k.
%\end{equation}
%The receiver can then decode $\bs_k$ directly from $\bz_k$ via a spatial demapper and demodulator. This paper is concerned with the joint 
%design of $\{\bF_\ell\}$ and $\{\bG_k\}$. 
Once the precoders are designed, the instantaneous sum rate of the system is 
\begin{equation}
R_{\rm sum} = \sum_{k=1}^K\log\left|\bI + 
\tilde{\bR}_k^{-1}\bH_{k,k}\bF_k\bF_k^*\bH_{k,k}^*\right|,
\label{eq:sum_rate}
\end{equation}
where
\begin{equation}
\tilde{\bR}_k = \bR_k + \sum_{\substack{\ell=1\\\ell\ne k}}^K \bH_{k,\ell}\bF_\ell\bF_\ell^*\bH_{k,\ell}^*
\end{equation}
is the interference plus noise covariance matrix at receiver $k$.
The instantaneous sum rate is an important metric for multiuser systems because of its ability to capture the total network throughput in a single scalar.
Notice that $R_{\rm sum}$ assumes ideal non-linear decoding of the signal. 
Although the proposed algorithms of Section~\ref{sec:algorithms} will design linear processing matrices that can form a part of a linear receiver design, 
they serve mainly to simplify the optimization and design of the precoders.
Design of high performance linear receivers is left to future work, except for the MMSE design presented in Section~\ref{subsec:mmse_opt}.
Thus for fair comparison, the sum rate equations assume an ideal decoding for all precoder designs.

Previous authors have shown that $KM/2$ spatial degrees of freedom are achievable in an $(M,M,K)$ interference channel 
Degrees of freedom $d$ is 
\begin{equation}
d = \lim_{{\rm SNR}\rightarrow\infty}\frac{C_{\rm sum}\left({\rm SNR}\right)}{\log {\rm SNR}},
\end{equation}
where $C_{\rm sum}$ is the sum capacity of the network, rather than the sum rate for our linear precoding model presented in~(\ref{eq:sum_rate}).

The key idea of interference alignment is to make $\sum_{\ell\ne k} S_\ell$ interferers appear as $N_k-S_k$ interferers at receiver $k$ for 
each $k$ by having them span a subspace of dimension $N_k-S_k$ of the $N_k$-dimensional receive space. 
Mathematically,
\begin{equation}
\sum_{\ell\ne k}\bH_{k,\ell}\bF_\ell = \sum_{i=1}^{N_k-S_k}a_i\bc_k^{(i)}, \forall k,
\end{equation}
where $\{\bc_k^{(i)}\}$ are basis vectors for the subspace at receiver $k$ in which all interference must lie.
Then receiver $k$ can then resolve its $S_k$ streams with a linear receiver interference-free~\cite{CadJaf:Interference-Alignment-and-Degrees:08}.
%Consider, for example, a (2,2,3) interference channel with 
%$S=1$:
%\begin{equation}
%\by_1 = \bH_{1,1}{\bf f}_1s_1 + \bH_{1,2}{\bf f}_2s_2 + \bH_{1,3}{\bf f}_3s_3 + \bv_1.
%\label{eq:y_1}
%\end{equation}
%We focus on receiver $k=1$, though the basic structure of the equations are identical for any $k$. 
%Suppose now that the precoders ${\bf f}_2$ and ${\bf f}_3$ are designed such that 
%$\bH_{1,2}{\bf f}_2 = a_1\bH_{1,3}{\bf f}_3$ with $a_1\in\mathbb{C}$. Then~(\ref{eq:y_1}) can be rewritten as
%\begin{equation}
%\by_1 = \bH_{1,1}{\bf f}_1s_1 + \bH_{1,3}{\bf f}_3\tilde{s}_{23} + \bv_1,
%\end{equation}
%where $\tilde{s}_{23} = s_3+a_1s_2$. This is mathematically equivalent to a multiple access scenario where two 
%transmitters attempt
%to send the scalars $s_1$ and $\tilde{s}_{23}$, respectively, to receiver 1. Since receiver 1 possesses two antennas, it can fully
%cancel $\tilde{s}_{23}$ and decode $s_1$ interference-free using a simple zero-forcing receiver.  
%For interference alignment to make sense and be useful, %having $\bH_{1,2}{\bf f}_2=\alpha_1\bH_{1,3}{\bf f}_3$ is not enough;
%the same general principle must be applied at every receiver. Thus, the following
%requirements must be satisfied:
%\begin{eqnarray}
%\bH_{1,2}{\bf f}_2 & = & a_1\bH_{1,3}{\bf f}_3\label{eq:ia_eqs_1}\\
%\bH_{2,1}{\bf f}_1 & = & a_2\bH_{2,3}{\bf f}_3\label{eq:ia_eqs_2}\\
%\bH_{3,1}{\bf f}_1 & = & a_3\bH_{3,2}{\bf f}_2.\label{eq:ia_eqs_3}
%\end{eqnarray}
For the three user interference channel it is possible to directly find closed-form solutions to $\{\bF_\ell\}$.
for any $S\le M/2$. Such solutions for obtaining $KM/2$ degrees of freedom, however, in the $(M,M,K)$ interference channel
with $K>3$ users do not appear to be possible. Closed-form solutions, even for a reduced multiplexing gain, are 
unknown~\cite{YetJafKay:Feasibility-Conditions-for-Interference:09} except in special cases~\cite{TreGuiRie:On-the-achievability-of-interference-alignment:09}.
A viable alternative for the general case are alternating minimizations.
The next section reviews the existing designs and proposes new algorithms for finding high-rate solutions at finite-SNR 
in the MIMO interference channel.
\section{Iterative Algorithms Via Alternating Minimization}\label{sec:algorithms}
This section presents iterative solutions for precoders in the MIMO interference channel using an alternating minimization to solve various optimization objectives.
%Ideally we aim to find MIMO precoders that maximize the sum rate of the network for any given channel realization. This requires a solution for
%the $\{\bF\ell\}$ terms in~(\ref{eq:sum_rate}), which is a difficult linear algebra problem because of the matrix inverse and co-dependence of variables.
This section proposes three new metrics which aim to approximate a sum rate maximization with better finite-SNR rates than previous work. 
%Section~\ref{sec:sims} compares the sum rate performance of each of the proposed designs in various environments through numerical simulations.

The algorithms presented may be implemented in a distributed or centralized manner similar to~\cite{PetHea:Interference-Alignment-Via-Alternating:09}.
These algorithms share a common structure. Each algorithm is designed to optimize a global objective
$\mathcal{J}$ that incorporates the performance of each data link in the network. The objective is a function of the precoders $\{\bF_\ell\}$,
the channels $\{\bH_{k,\ell}\}$ between all nodes\footnote{Some of the algorithms will not make use of the data links, 
instead focusing on minimizing post-processing interference.}, and a processing matrix at each receiver, the structure of which will
vary across designs\footnote{These matrices are not necessarily designed to function as spatial equalizers, instead serving mainly to 
simplify the design and optimization of the precoders. With the exception of our joint-MMSE algorithm, design of high performance linear receivers 
is left to future work.}. The free variables are the $K$ precoders and $K$ receive processing matrices.

A closed-form solution for a global optimization of any of the objectives in this section is unknown. We therefore turn to an 
alternating minimization\footnote{Alternating maximizations can be converted into alternating minimizations, 
so we focus on alternating minimization.} approach for the $2K$ variables~\cite{CsiTus:Information-geometry-and-alternating:84}. 
In general, an alternating minimization arbitrarily initializes $2K-1$ variables
and, assuming these variables are fixed, solves for the remaining one. It stores this solution, and moves to another variable, finding a
new solution for it assuming the rest of the variables are fixed. Each variable in turn is solved for during each iteration.  
Note that this procedure is convenient only if there is a simple or even closed-form solution for each of the variables assuming
the rest are fixed. Finally, for each of the designs, with the exception of the proposed maximum SINR design, the precoders
may be derived in parallel, since their solutions at any step of the algorithm do not depend on each other.
%The algorithms in this paper will find that each precoder is only a function of the receivers (not other precoders), and vice versa. 
%Thus, we only need to initialize the $K$ precoders (or $K$ receivers) and can solve for all $K$ receivers (precoders) simultaneously.

\subsection{Subspace Optimization}\label{sec:subspace_opt}
A direct algorithm for the interference channel inspired by interference alignment~\cite{CadJaf:Interference-Alignment-and-Degrees:08} 
is to precode the signal at transmitter $\ell$ such that the coordinated interference caused by transmitter $\ell$ at receiver $k\ne\ell$ is nearly orthogonal
to a subspace (with orthonormal basis ${\bf\Phi}_k$) of its receive 
space~\cite{GomCadJaf:Approaching-the-Capacity-of-Wireless:08,PetHea:Interference-Alignment-Via-Alternating:09}. 
This subspace is then jointly designed along with the precoders to optimize an appropriate cost function.
One way of performing this optimization is to minimize the total ``leakage interference''~\cite{GomCadJaf:Approaching-the-Capacity-of-Wireless:08}
that remains at each receiver after attempting to cancel the coordinated interference by left-multiplication with ${\bf\Phi}_k^*$ for each $k$.
The global function to optimize is thus
\begin{equation}
\mathcal{J}_{\rm IA} = \sum_{k=1}^{K}\mathbb{E}\left\|{\bf\Phi}_k^*\sum_{\substack{\ell=1\\ \ell\ne k}}^{K}
  \bH_{k,\ell}\bF_\ell\bs_\ell\right\|^2_F.
\end{equation}
The expectation in $\mathcal{J}_{\rm IA}$ and all subsequent analysis is over $\bs_k$ (and $\bv_k$ where applicable), $k\in\{1,\dots,K\}$.
Evaluating the expectation and exploiting independence of the signals,
\begin{equation}
\mathcal{J}_{\rm IA} = \sum_{k=1}^K\sum_{\substack{\ell=1\\ \ell\ne k}}^K\|{\bf\Phi}_k^*\bH_{k,\ell}\bF_\ell\|_F^2,
\label{eq:j_ia}
\end{equation}
which is termed ``interference leakage'' in~\cite{GomCadJaf:Approaching-the-Capacity-of-Wireless:08}.
The precoders $\{\bF_\ell\}$ are constrained to have mutually orthogonal columns with a per-stream power constraint so that 
$\bF_\ell^*\bF_\ell=\frac{\rho_\ell}{S_\ell}\bI,\forall\ell$. Although we could enforce a total power constraint on the precoders
(and, coincidentally in this case, get the same solution), orthogonality is desired in MIMO precoding designs to aid with
feedback of channel state~\cite{LovHeaSan:What-is-the-value-of-limited:04}. The receive subspace bases $\{{\bf\Phi}_k\}$ are orthonormal 
by definition so that ${\bf\Phi}_k^*{\bf\Phi}_k = \bI$.  
The objective is thus
\begin{eqnarray}
\mathrm{minimize} & \mathcal{J}_{\rm IA}\left(\{\bF_\ell\},\{{\bf\Phi}\}\right)\nonumber\\
\mathrm{subject~to} & \bF_\ell^*\bF_\ell=\frac{\rho_\ell}{S_\ell}\bI, \ell\in\{1,\dots,K\}\label{eq:ia_obj}\\
{} & {\bf\Phi}_k^*{\bf\Phi}_k=\bI, k\in\{1,\dots,K\}\nonumber.
\end{eqnarray}
The optimization~(\ref{eq:ia_obj}) is intuitively pleasing since, with perfect interference alignment, $\mathcal{J}_{\rm IA}=0$, and without interference
alignment, $\mathcal{J}_{\rm IA}>0$. That is, interference alignment, if possible, achieves the global minimum for this function. 

Deriving a closed-form solution to~(\ref{eq:ia_obj}) for $K>3$ users is difficult due to the inter-dependence of each precoder and 
receive interference-free subspace. 
A simple approach, which is guaranteed to converge, is to use an alternating minimization~\cite{CsiTus:Information-geometry-and-alternating:84}. 
%The algorithm can be first initialized with random $\{{\bf\Phi}_k\}$, 
%then solved for $\{\bF_\ell\}$ assuming $\{{\bf\Phi}_k\}$ are fixed, then solved for $\{{\bf\Phi}_k\}$ assuming $\{\bF_\ell\}$ are fixed, and
%so on. Note that $\mathcal{J}_{\rm IA}$ is rotationally invariant with respect to $\{\bF_\ell\}$ and $\{{\bf\Phi}_k\}$. Since each of these
%are also constrained to be truncated unitary matrices, then the solution space for the $\{\bF_\ell\}$ and $\{{\bf\Phi}_k\}$ lie on the
%Grassmann manifold and we can use existing manifold gradients to solve for the desired matrices at each 
%step~\cite{EdeAriSmi:The-Geometry-of-Algorithms-with:98}. 
The derivation of this solution is in~\cite{GomCadJaf:Approaching-the-Capacity-of-Wireless:08} and our previous work in~\cite{PetHea:Interference-Alignment-Via-Alternating:09} 
and is not included here for efficiency. At each step, the solution for each $\bF_\ell$ is
\begin{equation}
\bF_\ell = \nu_{\rm min}^{S_k}\left(\sum_{\substack{k=1\\ k\ne\ell}}^K\bH_{k,\ell}^*{\bf\Phi}_k{\bf\Phi}_k^*\bH_{k,\ell}\right),
\label{eq:subspace_precoder}
\end{equation}
and, with all precoders given, the solution for each ${\bf\Phi}_k$ is
\begin{equation}
{\bf\Phi}_k = \nu_{\rm min}^{S_k}\left(\sum_{\substack{\ell=1\\\ell\ne k}}^K\bH_{k,\ell}\bF_\ell\bF_\ell^*\bH_{k,\ell}^*\right).
\label{eq:subspace_decoder}
\end{equation}
To run the algorithm, arbitrary receive subspaces for each receiver are used for initialization and an arbitrary orthonormal basis 
${\bf\Phi}_k$ for each subspace is found.
This subspace is ideally reserved for user $k$'s signal, thus coordinated interference at receiver $k$ is ideally orthogonal to this subspace. 
Then, for each $\ell$, the algorithm finds the precoder matrix $\bF_\ell$ such
that total coordinated interference caused at each node (other than at node $\ell$) has maximum squared Euclidean distance between it and the
subspace spanned by the columns of each ${\bf\Phi}_k$ using~(\ref{eq:subspace_precoder}). Given these new precoders, the algorithm can update the
receive subspaces to be those that span the columns of the matrices with minimum sum squared Euclidean distance to the interference caused 
by the fixed 
precoders using~(\ref{eq:subspace_decoder})~\cite{PetHea:Interference-Alignment-Via-Alternating:09}. 
This can be carried out until $\mathcal{J}_{\rm IA}(t)<\epsilon$ if feasibility conditions are met, or 
$\mathcal{J}_{\rm IA}(t-1)-\mathcal{J}_{\rm IA}(t)<\epsilon$ otherwise, for an arbitrary convergence threshold $\epsilon$. 
%The rest of the algorithms presented here are generalizations of this one.

Note that each receiver must still separate the desired spatial streams after the coordinated interference has been 
canceled with left multiplication of ${\bf\Phi}^*$. Standard linear designs, such as zero forcing or MMSE, can be employed for this purpose.
%, instead of designing the receive spatial filters $\{\bG_k\}$, this algorithm designs interference-free subspaces, 
%whose column-wise orthonormal bases we %denote as $\{{\bf\Phi}_k\}$. 
%Thus, if the interference (at receiver $k$, for instance) is orthogonal to the subspace, it can be fully canceled by left-multiplication by ${\bf\Phi}_k^*$. 
Thus, the receiver can form a linear receive filter $\bG_k$ by multiplying ${\bf\Phi}_k$ and 
the linear spatial filter $\bW_k$, which neglects coordinated inter-user interference and equalizes only the desired signal, so that $\bG_k={\bf\Phi}_k\bW_k$.
Then the vector $\hat{\bs}_k=\bG_k^*\by_k$ is the interference-free estimate of the original transmitted vector $\bs_k$.

\subsection{Minimum Interference Plus Noise Leakage (INL)}\label{sec:colored_noise}
\begin{figure}
\centering
\includegraphics[width=3.5in]{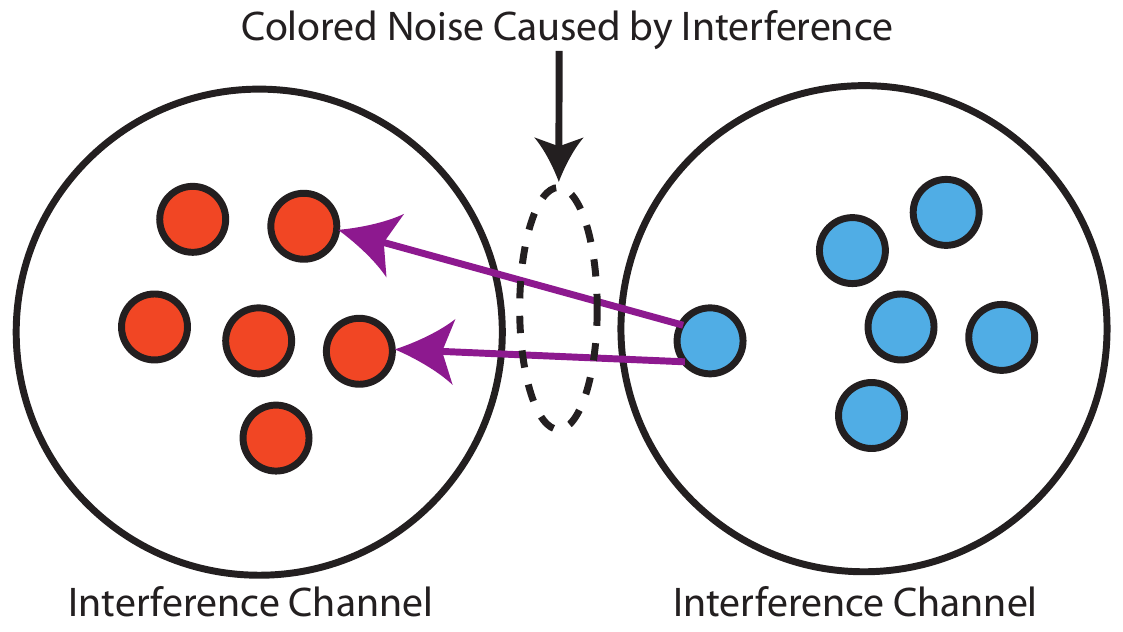}
\caption{Receivers modeled by an interference channel may experience uncoordinated interference, modeled as colored noise, from part of the network not modeled as
being a part of the same interference channel.}
\label{fig:colored_source}
\end{figure}
The subspace approach of~\cite{GomCadJaf:Approaching-the-Capacity-of-Wireless:08,PetHea:Interference-Alignment-Via-Alternating:09},
outlined in Section~\ref{sec:subspace_opt}, aims at aligning interference, which is capacity-optimal as the ratio of signal power to receiver noise
power tends to infinity.
If colored noise exists in any receiver, however, the IA subspaces might be chosen to align with the noise to cancel it as well as the
interference. Such colored noise may be due to an interference source outside of the coordinated portion of the network modeled as an interference
channel, as shown in Figure~\ref{fig:colored_source}. This interference is referred to as \emph{uncoordinated interference}. 
We therefore focus on algorithms that take noise into account in their optimization. 
Note that these approaches have a ``global'' objective function limited to the users cooperating in interference channel, such as inside a single cluster 
in Figure~\ref{fig:colored_source}, and thus assume the uncoordinated interferers of other clusters have fixed covariance over the optimization and transmission time.

The objective of the subspace algorithm of Section~\ref{sec:subspace_opt} is to minimize the total post-processing coordinated interference power, 
also known as \emph{interference leakage} or \emph{interference power} 
in~\cite{GomCadJaf:Approaching-the-Capacity-of-Wireless:08,GomCadJaf:Approaching-the-capacity-of-wireless:09}.
Thus, one intuitive solution is to minimize the total interference plus noise leakage, or INL.
Mathematically, this is represented with the global performance function
\begin{equation}
\mathcal{J}_{\rm INL}= \sum_{k=1}^{K}{\mathbb E}
\left\|{\bf\Phi}_k^*\left(\sum_{\substack{\ell=1\\ \ell\ne k}}^{K}\bH_{k,\ell}\bF_\ell\bs_\ell+\bv_k\right)\right\|^2_F,
\end{equation}
where $\bv_k$ is the received noise vector observed at receiver $k$. Expanding the expectation and exploiting the independence of the signal and noise vectors, 
the objective becomes
\begin{equation}
\mathcal{J}_{\rm INL}=\sum_{k=1}^K\sum_{\substack{\ell=1\\ \ell\ne k}}^K\left\|{\bf\Phi}_k^*\bH_{k,\ell}\bF_\ell\right\|^2_F + 
                      {\rm tr}\left({\bf\Phi}_k^*\bR_k{\bf\Phi}_k\right),
\label{eq:j_ian}
\end{equation}
where $\bR_k=\mathbb{E}\bv_k\bv_k^*$ is the covariance matrix of the noise at receiver $k$.
The objective is then
\begin{eqnarray}
\mathrm{minimize} & \mathcal{J}_{\rm INL}\left(\{\bF_\ell\},\{{\bf\Phi}_k\}\right)\nonumber\\
\mathrm{subject~to} & \bF_\ell^*\bF_\ell=\frac{\rho_\ell}{S_\ell}\bI, \ell\in\{1,\dots,K\}\label{eq:noise_obj}\\
{} & {\bf\Phi}_k^*{\bf\Phi}_k=\bI, k\in\{1,\dots,K\}\nonumber.
\end{eqnarray}
The constraints on the precoders and receive subspaces are identical to those in Section~\ref{sec:subspace_opt}.
Further, since $\mathcal{J}_{\rm INL}$ is rotation-invariant to
each of the variables, the solutions lie on the Grassmann manifold and techniques derived for it can be used.
Since $\|\bA\|_F^2={\rm tr}\left(\bA\bA^*\right)$, $\mathcal{J}_{\rm INL}$ can be rewritten as
\begin{equation}
\mathcal{J}_{\rm INL} = \sum_{k=1}^K\sum_{\substack{\ell=1\\\ell\ne k}}^K
    {\rm tr}\left({\bf\Phi}_k^*\left(\bH_{k,\ell}\bF_\ell\bF_\ell^*\bH_{k,\ell}^* + \bR_k\right){\bf\Phi}_k\right),
\end{equation}
which for fixed $\{\bF_\ell\}$ is minimized by~\cite{Lut:Handbook-of-Matrices:97}
\begin{equation}
{\bf\Phi}_k^{opt} = \nu_{\rm min}^{S_k}\left(\bR_k+\sum_{\substack{\ell=1\\\ell\ne k}}^K\bH_{k,\ell}\bF_\ell\bF_\ell^*
    \bH_{k,\ell}^*\right).
\end{equation}
For the precoders $\{\bF_\ell\}$, it is sufficient to note that, for fixed $\{{\bf\Phi}_k\}$, minimizing $\mathcal{J}_{\rm INL}$ with respect
to $\{\bF_\ell\}$ is equivalent to minimizing $\mathcal{J}_{\rm IA}$ with respect to $\{\bF_\ell\}$, as is seen by comparing
(\ref{eq:j_ian}) and (\ref{eq:j_ia}). Thus, the precoder solution is identical to~(\ref{eq:subspace_precoder}).
%As shown in Appendix~\ref{app:ia_noise}, the solution to the receive subspace matrix is
%\begin{equation}
%{\bf\Phi}_k = \nu_{\rm min}^{S_k}\left(\sum_{\substack{\ell=1\\\ell\ne k}}^K\bH_{k,\ell}\bF_\ell\bF_\ell^*\bH_{k,\ell}^* + \bR_k\right).
%\label{eq:noise_decoder}
%\end{equation}
%The precoder solutions are identical to that of (\ref{eq:subspace_precoder}).
This solution effectively tries to align the coordinated interference with the dominant directions of the noise (or uncoordinated interference)
if the noise has significant energy. In particular, if the noise is highly correlated spatially with a rank-one covariance matrix, then
$\bR_k=\sigma^2_k\ba_k\ba_k^*$ and this algorithm will attempt to align the interference to $\ba_k$ if possible. Such noise, which may
correspond to a single-stream uncoordinated interferer not part of the cooperating network, might then be mitigated, although
full removal is unlikely. 
We can also prove the following quantitative conclusions. 
\begin{proposition}
If $\bR_k=\sigma_k^2\bI$ $\forall k\in\{1,\dots,K\}$, then minimizing $\mathcal{J}_{\rm INL}$ is 
equivalent to minimizing $\mathcal{J}_{\rm IA}$.
\label{prop:diag_noise}
\end{proposition}
\begin{IEEEproof}
By definition,
\begin{eqnarray}
\mathcal{J}_{\rm INL} & = & \sum_{k=1}^K\left\|{\bf\Phi}_k^*\sum_{\ell\ne k}\bH_{k,\ell}\bF_\ell\right\|_F^2 + 
         {\rm tr}\left({\bf\Phi}_k^*\bR_k{\bf\Phi}_k\right)\nonumber\\
{} & = & \sum_{k=1}^K\left\|{\bf\Phi}_k^*\sum_{\ell\ne k}\bH_{k,\ell}\bF_\ell\right\|_F^2 + 
         \sigma_k^2{\rm tr}\left({\bf\Phi}_k^*{\bf\Phi}_k\right)\nonumber\\
{} & = & \mathcal{J}_{\rm IA} + \sum_{k=1}^K\sigma_k^2S_k.\label{eq:same_subspace}
\end{eqnarray}
Since the summation in~(\ref{eq:same_subspace}) is independent of any of the free variables, minimizing $\mathcal{J}_{\rm INL}$ is
equivalent to minimizing $\mathcal{J}_{\rm IA}$ when the noise is spatially white at each receiver.
\end{IEEEproof}
\begin{proposition}
As $\rho_k\rightarrow\infty$ or $\|\bR_k\|_F\rightarrow 0$ for all $k$, $\mathcal{J}_{\rm INL}$ converges to $\mathcal{J}_{IA}$. Thus,
the subspace algorithm with noise consideration has the same SNR scaling as the pure interference alignment algorithm.
\label{prop:no_noise}
\end{proposition}
\begin{IEEEproof}
Define $\lambda_k$ as the largest eigenvalue of Hermitian matrix $\bR_k$. Then,
\begin{eqnarray}
\mathcal{J}_{\rm INL} & = & \mathcal{J}_{\rm IA} + \sum_{k=1}^K{\rm tr}\left({\bf\Phi}_k^*\bR_k{\bf\Phi}_k\right)\nonumber\\
{} & \le & \mathcal{J}_{\rm IA} + \sum_{k=1}^K\lambda_kS_k.
\end{eqnarray}
For any arbitrary $\{\bR_k\}$, we define a sequence of functions 
\begin{equation}
\mathcal{J}_{\rm INL}^{(n)}\doteq \mathcal{J}_{\rm IA} + \frac{1}{n}\sum_{k=1}^K\lambda_kS_k,
\end{equation}
corresponding to a sequence of noise covariance matrices $\bR_k^{(n)} = \bR_k/n$, so that
$\|\bR_k\|_F\rightarrow 0$ as $n\rightarrow\infty$.
Then for any $\epsilon>0$, 
\begin{equation}
\left|\mathcal{J}_{\rm INL}^{(n)} - \mathcal{J}_{\rm IA}\right| \le \epsilon
\end{equation}
for all $n>\sum_{k=1}^K\lambda_kS_k/\epsilon$.
\end{IEEEproof}
From the proof, we also note that $\mathcal{J}_{\rm INL}\ge\mathcal{J}_{\rm IA}$ and $\min \mathcal{J}_{\rm INL}=0$ iff 
$\bR_k$ is singular for all $k$ and the columns of the interference-aligning receiver matrices $\{{\bf\Phi}_k\}$ lie in the null spaces of their
respective noise covariance matrices $\{\bR_k\}$.
The metric $\mathcal{J}_{\rm INL}$ is, in fact, likely to have a positive global minimum unless the total number of streams is reduced below the
degrees of freedom of the network, even if the noise is correlated, because the noise subspaces at different receivers will be not perfectly 
alignable almost surely. Adapting the number of streams in the network to improve finite-SNR performance is an interesting problem that is 
beyond the scope of this paper.

%If, however, the noise 
%is spatially white so that $\bR_k=\sigma^2\bI$, the solution to this algorithm
%will be identical to that of the previous subspace algorithm without noise consideration. This can easily be shown because $\bR_k$
%would not change the eigenvectors of the argument in~(\ref{eq:noise_decoder}), and it would thus reduce to~(\ref{eq:subspace_decoder}).
%In either case, the transmit precoder is identical to that in~(\ref{eq:subspace_precoder}). Further, as 
%$\rho_k\rightarrow\infty$, this algorithm again becomes the subspace algorithm in Section~\ref{sec:subspace_opt}. It thus has the same 
%SNR scaling as interference alignment.

In an idealized system with 
Gaussian signaling, colored noise may correspond to uncoordinated interference from outside the network of interest. For instance,
consider the scenario of a cellular network across a metropolitan area. The strategy for this network may be to coordinate three adjacent sectors
to use interference alignment (via subspace optimization) to transmit to one mobile per sector in the downlink. For a regular IA solution,
the uncoordinated interference arriving at each receiver from sectors outside the coordination area would be ignored or modeled as spatially white. 
The min-INL algorithm would be able to exploit the knowledge of this uncoordinated interference and account for it as necessary.

The algorithms of Sections~\ref{sec:subspace_opt} and~\ref{sec:colored_noise} aim to align the coordinated interference, which in turn maximizes capacity
in a fully connected high-SNR network. We have seen in Proposition~\ref{prop:diag_noise} that in finite-SNR environments, white Gaussian noise 
does not change the solutions of subspace methods. Although this section has presented an approach for networks with colored noise, 
algorithms with better throughput performance in finite-SNR regimes are desired, especially since most networks are not likely to be fully
connected and thus may operate with low interference-to-noise ratio (INR), where subspace methods are not likely optimal even with colored noise considerations. 
%effectively minimizes the post-processing interference power,  
%making the signal-to-interference ratio (SIR) infinite if perfect alignment is achieved. If noise is present in the receiver, 
%however, it will become the limiting factor in
%the ability to decode the signal. The subspace optimization algorithm ignores noise and therefore may result in a low post-processing 
%signal-to-noise ratio (SNR). 
To illustrate the problem of implementing subspace algorithms in a real network, consider the following argument.
Suppose all interfering links $\{\bH_{k,\ell}\}, k\ne\ell$ have a path loss coefficient $\beta$ whereas direct links
have a path loss coefficient of 1. The subspace precoder design will then not depend at all on the value of $\beta$ since the scalar
multiplication does not change the direction of the signal. If the receivers use their interference suppression filters $\{\bU_k\}$
to cancel the interference, then the throughput of the system will be independent of $\beta$. Thus, subspace algorithms treat weak and
strong interferers equally, without exploiting the possible capacity gains available when interference is weakened.
If no noise exists in the system, this is perfectly fine, since the receiver will still have an interference-free signal that it could decode
perfectly. Realistically, however, a dynamic network would benefit from adapting its behavior to the relative interference energy. As shown numerically in Section~\ref{sec:sims},
the algorithms proposed in Sections~\ref{subsec:mmse_opt} and~\ref{subsec:max_sinr} are more suited to such adaptation than the subspace method of Section~\ref{sec:subspace_opt}.

%%%%%%%%%%%%%%%%%%%%%%%%%%%%%%%%%%%%%%%%%%%%%%%%%%%%%%%%%%%%%%%%%%%%
\subsection{Mean Squared Error Minimization}\label{subsec:mmse_opt}
A common metric for accounting for noise in linear receivers in wireless communication systems is the mean squared error. For example,
a zero-forcing linear MIMO receiver simply inverts the channel, and results in coloring and amplification of noise. An MMSE receiver
balances the effects of noise with that of inverting the channel depending on the relative energy of each.
This same concept can be applied to interference alignment, where the transmitter and receiver balance their wish to align the coordinated interference
with the need for keeping the signal level well above the noise. 

Joint MMSE designs for MIMO channels have been studied for years and have been applied to the point-to-point
model~\cite{Sal:Digital-transmission-over:85,YanRoy:On-joint-transmitter-and-receiver:94,SamPau:Joint-transmit-and-receive:99} 
and the broadcast channel~\cite{TenAdv:Joint-multiuser-transmit-receive:04,ZhaWuZho:Joint-linear-transmitter:05}. 
The development for the interference channel is distinguished from previous work
in that precoders and receivers need to be designed for multiple transmitters and receivers, rather than just the multiple transmitters 
\emph{or} receivers as in the multi-user case, or a single transmitter and receiver in the point-to-point case. 

As opposed to objectives discussed in Sections~\ref{sec:subspace_opt} and \ref{sec:colored_noise}, the MMSE directly designs the receive
spatial filters $\{\bG_k\}$. That is, the output of the product $\bG^*_k\by_k$ is the estimate $\hat{\bs}_k$ of $\bs_k$, and the MMSE criterion
minimizes the expected sum of the norms between each $\hat{\bs}_k$ and $\bs_k$ for all $k$, yielding the objective
\begin{equation}
\mathcal{J}_{\rm MSE} = \sum_{k=1}^K{\mathbb E}\|\bG_k^*\by_k - \bs_k\|^2.
\end{equation}
Substituting~(\ref{eq:y_k2}) for $\by_k$ results in a global performance function of
\begin{equation}
\mathcal{J}_{\rm MSE} = \sum_{k=1}^{K}{\mathbb E}
\biggl\|\bG_k^*\biggl(\bH_{k,k}\bF_k\bs_k+\sum_{\substack{\ell=1\\ \ell\ne k}}^{K}\bH_{k,\ell}\bF_\ell\bs_\ell+\bv_k\biggr)
-\bs_k\biggr\|^2_F,
\end{equation}
and an optimization objective of
\begin{eqnarray}
\mathrm{minimize} & \mathcal{J}_{\rm MSE}\left(\{\bF_\ell\},\{\bG_k\}\right)\nonumber\\
\mathrm{subject~to} & \|\bF_\ell\|_F^2\le \rho_\ell, \ell\in\{1,\dots,K\}.
\label{eq:mmse_obj}
\end{eqnarray}
Expanding the expectation and simplifying, the optimization is equivalent to
\begin{eqnarray}
\mathrm{minimize} & \sum_{k=1}^K 
    {\rm tr}\left(\bG_k^*\left(\tilde{\bR}_k+\bH_{k,k}\bF_k\bF_k^*\bH_{k,k}^*\right)\bG_k\right)\nonumber\\
{} & -2\mathbb{R}\left\{{\rm tr}\left(\bG_k^*\bH_{k,k}\bF_k\right)\right\}\nonumber\\
\mathrm{subject~to} & \|\bF_\ell\|_F^2\le \rho_\ell, \ell\in\{1,\dots,K\}.
\label{eq:j_mmse}
\end{eqnarray}

In general, MMSE solutions with an orthogonality constraint are more difficult to derive. Thus, we relax the orthogonality
constraint to a total power inequality constraint $\|\bF_\ell\|_F^2\le \rho_\ell, \forall\ell$, and resort to a solution satisfying the 
Karush-Kuhn-Tucker (KKT) conditions as in previous joint MMSE solutions for different channel models~\cite{UluYen:Iterative-transmitter-and-receiver:04}. 
As shown in Appendix~\ref{app:mmse}, at each step the optimal precoders are
\begin{equation}
\bF_\ell = \left(\mu_\ell\bI + \sum_{k=1}^K\bH_{k,\ell}^*\bG_k\bG_k^*\bH_{k,\ell}\right)^{-1}\bH_{\ell,\ell}^*\bG_\ell,
\label{eq:mmse_f}
\end{equation}
where $\mu_\ell$ is the Lagrangian multiplier chosen to meet the power constraint. This may require a simple optimization 
(detailed in Appendix~\ref{app:mmse}) and has no known closed form. The optimal receivers are
\begin{equation}
\bG_k = \left(\sum_{\ell=1}^K\bH_{k,\ell}\bF_\ell\bF_\ell^*\bH_{k,\ell}^* + \bR_k\right)^{-1}\bH_{k,k}\bF_k,
\label{eq:mmse_g}
\end{equation}
where no further optimization needs to be performed because there is no constraint on the receiver.
As the following proposition shows, this design can be viewed as a generalization of previous designs for the point-to-point case. 
\begin{proposition}
With $\bH_{k,\ell}={\bf 0}$ for all $\ell, k$ such that $k\ne\ell$, (\ref{eq:mmse_f}) and (\ref{eq:mmse_g}) are equivalent to an MMSE design for
a point-to-point scenario. Further, as $\rho_k\rightarrow\infty$, the precoders and receivers diagonalize their respective information
links.
\end{proposition}
\begin{IEEEproof}
This is proven by substituting ${\bf 0}$ for each $\bH_{k,\ell}$, $k\ne\ell$, and referring to previous point-to-point 
results~\cite{YanRoy:On-joint-transmitter-and-receiver:94,SamPau:Joint-transmit-and-receive:99}.
\end{IEEEproof}
We also note that at high SNR and no coordinated inter-user interference, the MMSE algorithm will converge with one step, since
any initialization precoder $\bF_\ell$ is a fixed point of the algorithm and will minimize the MSE.

%Note that, unlike the previous algorithms, the output of the MMSE receiver is the actual estimate of $\bs_k$. The subspace optimization methods
%will require a conventional MIMO receiver with $\bG_k^*\bH_{k,k}\bF_k$ as the effective channel matrix to find an estimate for $\bs_k$.

The MMSE design is unique among those discussed in this paper. 
%For one, it does not have the reciprocity property. 
As discussed before, the MMSE receiver gives a direct estimate of $\bs_k$, while the others require a conventional
MIMO receiver after $\bG_k$ is applied. The MMSE receiver solution for fixed $\bF_\ell, \ell\in\{1,\dots,K\},$ is simply the conventional
MMSE MIMO receiver with colored noise. Further, the solution at each step is not in closed form, as an optimization needs to be 
done to meet the power constraint for the precoders. Lastly, the precoder solution is not orthogonal (or, conversely, a solution 
with orthogonal constraints is difficult to find). This algorithm may be difficult to implement because of these properties.

Finally, the min-INL optimization is equivalent to an MMSE problem that compares the post-processing output
to ${\bf\Phi}_k^*\bH_{k,k}\bF_k\bs_k$ instead of simply $\bs_k$. That is, 
\begin{equation}
\mathcal{J}_{\rm INL} = \sum_{k=1}^K\mathbb{E}\left\|{\bf\Phi}_k^*\by_k - {\bf\Phi}_k^*\bH_{k,k}\bF_k\bs_k\right\|_F^2.
\end{equation}
Thus, the receiver ${\bf\Phi}_k$ is expected to remove only the effects of coordinated interference
and white noise instead of having to correct for distortion created by the channel as well. This output must then be sent to a 
MIMO equalizer to remove inter-stream interference before symbol-by-symbol demodulation. 

\subsection{Signal-to-Interference-Plus-Noise-Ratio Maximization}\label{subsec:max_sinr}
The original subspace algorithm presented in Section~\ref{sec:subspace_opt} minimizes post-processing coordinated interference energy. 
The min-INL algorithm in Section~\ref{sec:colored_noise} adds consideration for noise leakage as well, which can improve performance under colored noise. 
The MMSE solution in Section~\ref{subsec:mmse_opt} indirectly accounts for signal
power by attempting to force the received signal to look like the intended signal before precoding and transmission. It is clear, however, that
a more desirable metric for maximizing the sum throughput would directly account for the post-processing signal-to-interference-plus-noise ratio (SINR).

%A more direct way
%to consider the signal power would be to maximize a global signal-to-interference-plus-noise ratio (SINR). Even under this narrow consideration,
This section presents an algorithm for maximizing total SINR in the network.
The optimization we use is not the only one that could be considered ``maximum SINR'', however, since total SINR for multiple nodes is not 
strictly defined in the literature. One may construct any number of global SINR metrics. 
%For instance, one may consider the sum of the SINRs at each receiver, the minimum
%SINR at each receiver, or the product of SINRs at each receiver. Any particular construct would have its own advantages and disadvantages,
%but one property it must have for our iterative algorithmic approach to work is that it must have a simple solution for each 
%step of the algorithm. 
%%For this reason we choose the global SINR metric to be the ratio of the sum of signal powers to the sum of interference-plus-noise powers. 
%We may be tempted to use the following global performance (SINR-like) function
%\begin{equation}
%\mathcal{L} = \frac{\sum_{k=1}^{K}\mathbb{E}\left\|\bG_k^*\bH_{k,k}\bF_k\bs_k\right\|_F^2}
%{\sum_{k=1}^K\mathbb{E}\left\|\bG_k^*\left(\sum_{\substack{\ell=1\\ \ell\ne k}}^{K}\bH_{k,\ell}\bF_\ell\bs_\ell+\bv_k\right)\right\|_F^2},
%\end{equation}
%since it is the ratio of the expected sum signal power to the expected sum interuser interference plus noise power over the network. 
%%Such a ratio is an intuitively pleasing aggregation of the networks signal power and noise power to approximate the sum throughput of the network. 
%Unfortunately, finding solutions for $\mathcal{L}$ is a difficult 
%problem~\cite{SheLiBro:A-convex-programming-approach:07} if $S_k>1$ for any $k$. 
Previous authors have considered the inter-stream interference for each transmit/receive pair and solved for the
precoding and receiver matrices one column at a time~\cite{GomCadJaf:Approaching-the-Capacity-of-Wireless:08}, resulting in 
non-orthogonal precoders and receive spatial filters, as in the MMSE case.  
%The maximum SINR algorithm in~\cite{GomCadJaf:Approaching-the-Capacity-of-Wireless:08}, however, solves for the equalizer columns by 
%maximizing the SINR for \emph{each particular stream}, 
%and solves for the precoder columns by maximizing the virtual SINR (the desired signal power divided by the interference
%caused by the transmitter) for each stream. Unlike the algorithms presented here, and the iterative IA procedure 
%in~\cite{GomCadJaf:Approaching-the-Capacity-of-Wireless:08}, 
That approach, however, is not an alternating optimization of a global objective function, and its convergence is unproven. 
We therefore reformulate the problem into a maximization of the sum signal power across the network divided by the sum interference power, 
incorporating the inter-stream interference for each user. 
The performance function becomes
\begin{equation}
\mathcal{J}_{\rm SINR} = \frac{\sum_{k=1}^{K}\sum_{n=1}^{S_k}\mathbb{E}\bigl|\bg_k^{(n)*}\bH_{k,k}{\bf f}_k^{(n)}s_k^n\bigr|^2}
{\sum_{k=1}^K\sum_{n=1}^{S_k}\mathbb{E}\bigl|\bg_k^{(n)*}\bigl(\sum_{\substack{\ell=1\\ \ell\ne k}}^{K}
\br_{k,\ell}+\br_{k,k}^{(n)}
+\bv_k\bigr)\bigr|^2},
\label{eq:j_sinr}
\end{equation}
where $\bg_k^{(n)}$ is the $n$th column of matrix $\bG_k$, 
\begin{equation}
\br_{k,\ell}=\sum_{m=1}^{S_\ell}\bH_{k,\ell}{\bf f}_\ell^{(m)}s_\ell^m
\end{equation}
is the pre-processing interference at receiver $k$ from transmitter $\ell$,
\begin{equation}
\br_{k,k}^{(n)}=\sum_{\substack{w=1\\ w\ne n}}^{S_k}\bH_{k,k}{\bf f}_k^{(w)}s_k^w
\end{equation}
is the pre-processing self-interference from streams $w\ne n$ at receiver $k$,
and $s_k^n$ is the $n$th entry of vector $\bs_k$.
Notice that
\begin{eqnarray}
R_{\rm sum} & = & \sum_{i=1}^K\sum_{n=1}^{S_i}\log\left(1+\frac{P_i^{(n)}}{I_i^{(n)} + N_i^{(n)}}\right)\nonumber\\
{} & \ge & \log\left(1+\sum_{i=1}^K\sum_{n=1}^{S_i}\frac{P_i^{(n)}}{I_i^{(n)} + N_i^{(n)}}\right)\nonumber\\
{} & \ge & \log\left(1+\frac{\sum_{i=1}^K\sum_{n=1}^{S_i}P_i^{(n)}}{\sum_{i=1}^K\sum_{n=1}^{S_i}I_i^{(n)} + N_i^{(n)}}\right)\nonumber\\
{} & = & \log\left(1+\mathcal{J}_{\rm SINR}\right)\nonumber,
\end{eqnarray}
where $P_i^{(n)}$ is the post-processing signal energy of the $n$th stream at the $i$th receiver, $I_i^{(n)}$ is the post-processing
interference energy, and $N_i^{(n)}$ is the post-processing noise energy seen by the stream.
The new objective~(\ref{eq:j_sinr}) is the sum of signal power in the network divided by the sum coordinated inter-user interference power
and inter-stream interference power after processing. 
By maximizing this ratio the algorithm can design the precoders to either decrease 
post-processing interference (the denominator) or increase signal power (the numerator) to improve total network performance.

The function $\mathcal{J}_{\rm SINR}$ is a generalized Rayleigh quotient and can be solved using generalized eigen decomposition.
%Expanding the expectation, the function is
%\begin{equation}
%\mathcal{J}_{\rm SINR} = \frac{\sum_{k=1}^K\sum_{n=1}^{S_k}\left|\bg_k^{(n)*}\bH_{k,k}{\bf f}_k^{(n)}\right|^2}
%  {\sum_{k=1}^K\sum_{n=1}^{S_k}\left(\sum_{\substack{\ell=1\\\ell\ne k}}^K\sum_{m=1}^{S_\ell}
%    \left|\bg_k^{(n)*}\bH_{k,\ell}{\bf f}_\ell^{(m)}\right|^2
%    + \sum_{\substack{w=1\\w\ne n}}^{S_k}\left|\bg_k^{(n)*}\bH_{k,k}{\bf f}_k^{(w)}\right|^2 + \bg_k^{(n)*}\bR_k\bg_k^{(n)}\right)},
%\label{eq:j_sinr}
%\end{equation}
and the optimization problem is 
\begin{eqnarray}
\mathrm{maximize} & \mathcal{J}_{\rm SINR}\left(\{{\bf f}_\ell^{(n)}\},\{\bg_k^{(n)}\}\right)\nonumber\\ 
\mathrm{subject~to} & \|{\bf f}_\ell^{(n)}\|^2 = \frac{\rho_\ell}{S_\ell},\forall n,\ell.
\label{eq:sinr_obj}
\end{eqnarray}
For tractability we constrain each stream's precoder to have an norm equality constraint so that $\|\bF_\ell\|^2_F=\rho_\ell$. For a larger
objective function, 
and increased complexity, we could also introduce an inequality constraint on each column and vary the transmit power over the streams. 
%In general, generalized eigenvalues will not be orthogonal so we again relax our constraint on the precoders to that of total energy. 
As shown in Appendix~\ref{app:sinr}, the solutions to the columns of the precoders are
\begin{eqnarray}
{\bf f}_\ell^{(n)} = \sqrt{\frac{\rho_\ell}{S_\ell}}\nu_{\rm max}\biggl(
  \biggl(~S_\ell q_\ell^{(n)}\bI + 
  \sum_{\substack{w=1\\ w\ne n}}^{S_\ell}\bH_{\ell,\ell}^*\bg_\ell^{(w)}\bg_\ell^{(w)*}\bH_{\ell,\ell}\nonumber\\ 
  +\sum_{\substack{k=1\\ k\ne\ell}}^K\sum_{m=1}^{S_k}\bH_{k,\ell}^*\bg_k^{(m)}\bg_k^{(m)*}\bH_{k,\ell}\biggr)^{-1}\nonumber\\
  \left(\bH_{\ell,\ell}^*\bg_\ell^{(n)}\bg_\ell^{(n)*}\bH_{\ell,\ell} + S_\ell r_\ell^{(n)}\bI\biggr)
  \right),
  \label{eq:sinr_precoders}
\end{eqnarray}
where $q_\ell^{(n)}$ is the sum of the terms in the denominator 
of~(\ref{eq:j_sinr}) that do not directly involve ${\bf f}_\ell^{(n)}$, and $r_\ell^{(n)}$ is the sum of the terms in the numerator
of~(\ref{eq:j_sinr}) that do not directly involve ${\bf f}_\ell^{(n)}$. The solutions to the columns of the receivers are
\begin{eqnarray}
\bg_k^{(n)} = \nu_{\rm max}\Biggl(\Biggl(~\hat{q}_k^{(n)}\bI + 
  \sum_{\substack{w=1\\ w\ne n}}^{S_k}\bH_{k,k}{\bf f}_k^{(w)}{\bf f}_k^{(w)*}\bH_{k,k}^* +\nonumber\\
  \sum_{\substack{\ell=1\\ \ell\ne k}}^K\sum_{m=1}^{S_\ell}\bH_{k,\ell}{\bf f}_\ell^{(m)}{\bf f}_\ell^{(m)*}\bH_{k,\ell}^*\Biggr)^{-1}\nonumber\\
  \left(\bH_{k,k}{\bf f}_k^{(n)}{\bf f}_k^{(n)*}\bH_{k,k}^* + 
  \hat{r}_k^{(n)}\bI\right)\Biggr),
  \label{eq:sinr_receivers}
\end{eqnarray}
where $\tilde{\nu}_{\rm max}\left(\bA,\bB\right)$ is the generalized eigenvector corresponding to the 
largest generalized eigenvalue of the matrix pair $\left(\bA,\bB\right)$, 
$\hat{q}_k^{(n)}$ and $\hat{r}_k^{(n)}$ are defined similarly as~(\ref{eq:sinr_precoders}) but with respect to $\bg_k^{(n)}$ instead
of ${\bf f}_\ell^{(n)}$.

With all other variables fixed, the solutions in~(\ref{eq:sinr_precoders}) and~(\ref{eq:sinr_receivers}) maximize the global SINR 
function~(\ref{eq:j_sinr}), whereas the solutions in~\cite{GomCadJaf:Approaching-the-Capacity-of-Wireless:08} give a suboptimal approximation
to this solution. As shown in Section~\ref{sec:sims}, this does not imply that an iterative algorithm using the proposed solutions will converge to
a larger objective than that of~\cite{GomCadJaf:Approaching-the-Capacity-of-Wireless:08}. For any given channel realization and initialization,
the two algorithms may give an identical result, or either may outperform the other. The simulation results in Section~\ref{sec:sims}
suggest, however, that the two algorithms perform similarly on average. Since the proposed design requires more network knowledge than that
of~\cite{GomCadJaf:Approaching-the-Capacity-of-Wireless:08}, the latter is more attractive for implementation. If the extra network
state knowledge is available, however, an intelligent design would be to run both algorithms and choose the design that works best for
each channel realization, resulting in a sum throughput higher than either algorithm could produce individually.
%The former, however, is
%mathematically proven to converge since it is an alternating optimization of the global objective function~(\ref{eq:j_sinr}).

Note that the IA algorithm will minimize the left-hand term of the denominator in $\mathcal{J}_{\rm SINR}$, 
and the min-INL algorithm will minimize the entire denominator (minus inter-stream interference). Certainly, with no noise (or,
more rigorously, as total signal to noise ratio goes to infinity), the two solutions are equivalent since maximizing the SINR will
reduce to maximizing the SIR, which, as discussed before, IA does. This fact was proven in~\cite{GomCadJaf:Approaching-the-capacity-of-wireless:09}.
%Other similarities between the two algorithms include reciprocity 
%(the optimal receiver would be the optimal transmitter in the reciprocal direction and vice versa). 

\subsection{Convergence and Initialization}\label{subsec:discuss}
This section analyzes some important details of the algorithms proposed in this paper. In particular, the focus is on variable initialization,
algorithm convergence, method of execution, obtainment of channel state, and precoder constraints. %Numerical results are presented in Section~\ref{sec:sims}.
%\subsection{Initialization and Convergence}\label{subsec:inits}
\begin{figure}
\centering
\includegraphics[width=3.5in]{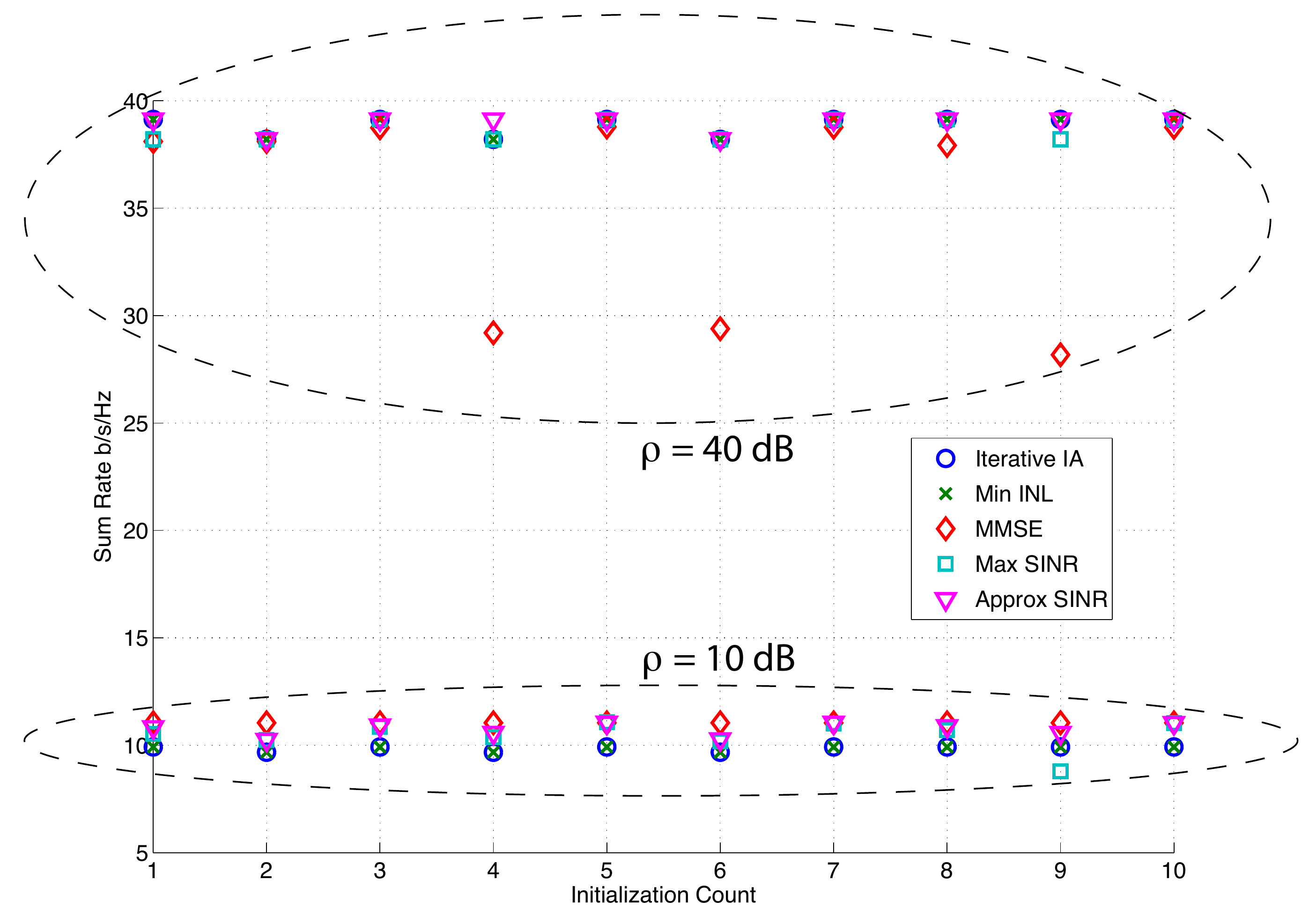}
\caption{Sum rate vs. initialization for each algorithm discussed in Section~\ref{sec:algorithms} run on a single channel realization of the $(2,2,3)$ MIMO IC
at $10$ and $40$ dB. Although the MMSE algorithm varies the most for this channel realization, this is not a general trend.}
\label{fig:init}
\end{figure}
We have found heuristically that arriving at a globally optimum point for the minimization algorithms (global optimality cannot be identified with the
max SINR algorithm) is highly likely even when initializing the precoders to truncated identity matrices; 
the throughput, however, is the real objective we wish to optimize, and these algorithms only approximate that optimization. 
Thus, different initializations of an algorithm may result in drastically different throughputs, even if they result in the same
final objective (or cost) function. For example, consider Figure~\ref{fig:init}. Each of the algorithms discussed in Section~\ref{sec:algorithms} 
was run on a fixed channel with 10 different random precoder initializations for the (2,2,3) MIMO IC at $\rho=40$ dB and $\rho=10$ dB. 
For $\rho=40$ dB, the MMSE algorithm varied most between
different initializations, but this is not indicative of the algorithms on the whole, just the behavior for this particular channel.
It appears that finding ``good''
initializations is not difficult; experimentation has shown that random initializations give as good of rates in these
algorithms as any ``intelligent'' initialization tried. If possible, multiple runs of the
algorithm should be made with different initializations for the best performance in terms of \emph{throughput}, as shown in Figure~\ref{fig:init}. 

Each of the algorithms from Section~\ref{sec:algorithms} are guaranteed to converge because the objectives are bounded 
and at each step are moving 
monotonically in the direction of that bound. Convergence to a global optimum is not guaranteed except when the 
objective has certain convexity-like properties~\cite{CsiTus:Information-geometry-and-alternating:84} that these algorithms 
are not proven to possess. Also, convergence of the cost function does not automatically imply convergence of the precoder designs, the analysis of
which is beyond our scope.

\section{Simulations}\label{sec:sims}
This section presents simulations of the algorithms presented in Section~\ref{sec:algorithms} to substantiate our claims and show that 
each of the algorithms can outperform the others in different regimes since none explicitly maximizes throughput. 
All of the simulations evaluate the expected sum rate with i.i.d. zero-mean unit-variance complex Gaussian coefficients for each channel, 
with the precoders for
each realization calculated with perfect CSI and as if the realization was flat in time and frequency. 
More realistic channel scenarios are considered in our related work~\cite{ElAPetHea:A-Study-of-the-Practicality-of-Interference:09}. 
Transmitter $k$ is assigned a deterministic transmit power $\rho_k$ and the link from transmitter $\ell$ to receiver $k$ 
has a deterministic path loss coefficient $\alpha_{k,\ell}$. Whereas in preceding analysis $\alpha_{k,\ell}$ was absorbed into $\bH_{k,\ell}$, 
in this section we pull it out for exposition. 
We also define $\gamma_{k,\ell}=\alpha_{k,\ell}\rho_{k,\ell}$ to be the expected SNR at receiver $k$ from transmitter $\ell$.
Thus, the sum rate is
\begin{equation}
R_{\rm sum} = \mathbb{E}_{\{\bH_{k,\ell}\}}\left\{\sum_{k=1}^K\log\left|\bI + 
\hat{\bR}_k^{-1}\alpha_{k,k}\bH_{k,k}\bF_k\bF_k^*\bH_{k,k}^*\right|\right\},
\label{eq:sims_rate}
\end{equation}
where
\begin{equation}
\hat{\bR}_k=\bR_k + \sum_{\ell\ne k}\alpha_{k,\ell}\bH_{k,\ell}\bF_\ell\bF_\ell^*\bH_{k,\ell}^*
\end{equation}
is the interference plus noise covariance.
Precoders are initialized randomly with orthonormal columns, as discussed in Section~\ref{subsec:discuss}, and each algorithm is presented with identical
initializations. Five random initializations are used for each channel realization, as motivated in Figure~\ref{fig:init}, and the initialization
that maximizes~(\ref{eq:sims_rate}) is kept while the others are thrown away.
In each plot presented in this section, $R_{\rm sum}$ is computed via Monte Carlo simulations using 1000 independent channel realizations.
Each iterative algorithm is run with 100 iterations each.

Each algorithm from Section~\ref{sec:algorithms} is compared with a random precoding scenario where
each precoder $\bF_\ell$ is chosen as the left singular vectors of a random Gaussian matrix, to enforce an orthogonality constraint.
That is, by \emph{Random Beamforming}, we mean that
\begin{equation}
\bF_\ell = \sqrt{\frac{\rho_\ell}{S_\ell}}\bU_\ell^{(S_\ell)},
\end{equation}
where $\bU_\ell^{(S_\ell)}$ are the first $S_\ell$ columns of the left singular matrix of a random matrix with i.i.d. zero-mean unit-variance
complex Gaussian coefficients.
A greedy approach is also included to show the benefit of cooperation in the MIMO interference 
channel~\cite{YeBlu:Optimized-signaling-for-MIMO:03,RosUluYat:Wireless-systems-and-interference:02}.  
In this design, each precoder $\bF_k$, $k\ne\ell$, is held fixed when designing $\bF_\ell$. Then 
\begin{equation}
\bU_\ell{\bf \Sigma}_\ell\bV_\ell^*=\biggl(\bR_\ell+\sum_{k\ne\ell}\alpha_{\ell,k}\bH_{\ell,k}\bF_k\bF_k^*\bH_{\ell,k}^*\biggr)^{-1/2}\sqrt{\alpha_{\ell,\ell}}\bH_{\ell,\ell},
\end{equation}
and
\begin{equation}
\bF_\ell = \sqrt{\frac{\rho_\ell}{S_\ell}}\bV_\ell^{(S_\ell)}.
\end{equation}
The greedy algorithm is not guaranteed to converge since it is not optimizing a global function, but it requires less channel estimation.
Finally, when the $K=3$ user interference channel is considered, the closed-form solution from~\cite{CadJaf:Interference-Alignment-and-Degrees:08} 
is also used for a baseline comparison.

We first introduce colored noise into the interference channel via an uncoordinated rank-one interferer in the network, as discussed in Section~\ref{sec:colored_noise}.
Defining $\bH_{k,E}$ as the MIMO channel from the uncoordinated rank-one interferer to receiver $k$ in the
interference channel, then receiver $k$ observes
\begin{equation}
\by_k = \sum_{\ell=1}^K\sqrt{\alpha_{k,\ell}}\bH_{k,\ell}\bF_\ell\bs_\ell + \sqrt{\alpha_{k,E}}\bH_{k,E}{\bf f}_Es_E + \bv_k,
\end{equation}
where since the uncoordinated interferer is rank-one, it is sending a single stream $s_E$ precoded with vector ${\bf f}_E$. Each receiver sees spatially white additive noise
on top of the signal and interference (coordinated and uncoordinated). Pure interference
alignment will ignore the uncoordinated interference, implicitly assuming it is spatially white. The rest of the algorithms will 
take the uncoordinated interferer into account but will not be able to fully suppress it without reducing the number of streams in the network
since the uncoordinated interferer, which is scaling its power with the transmitters inside the network, is reducing the degrees of freedom 
of the network, making it interference-limited. 

\begin{figure}
\centering
\includegraphics[width=3.5in]{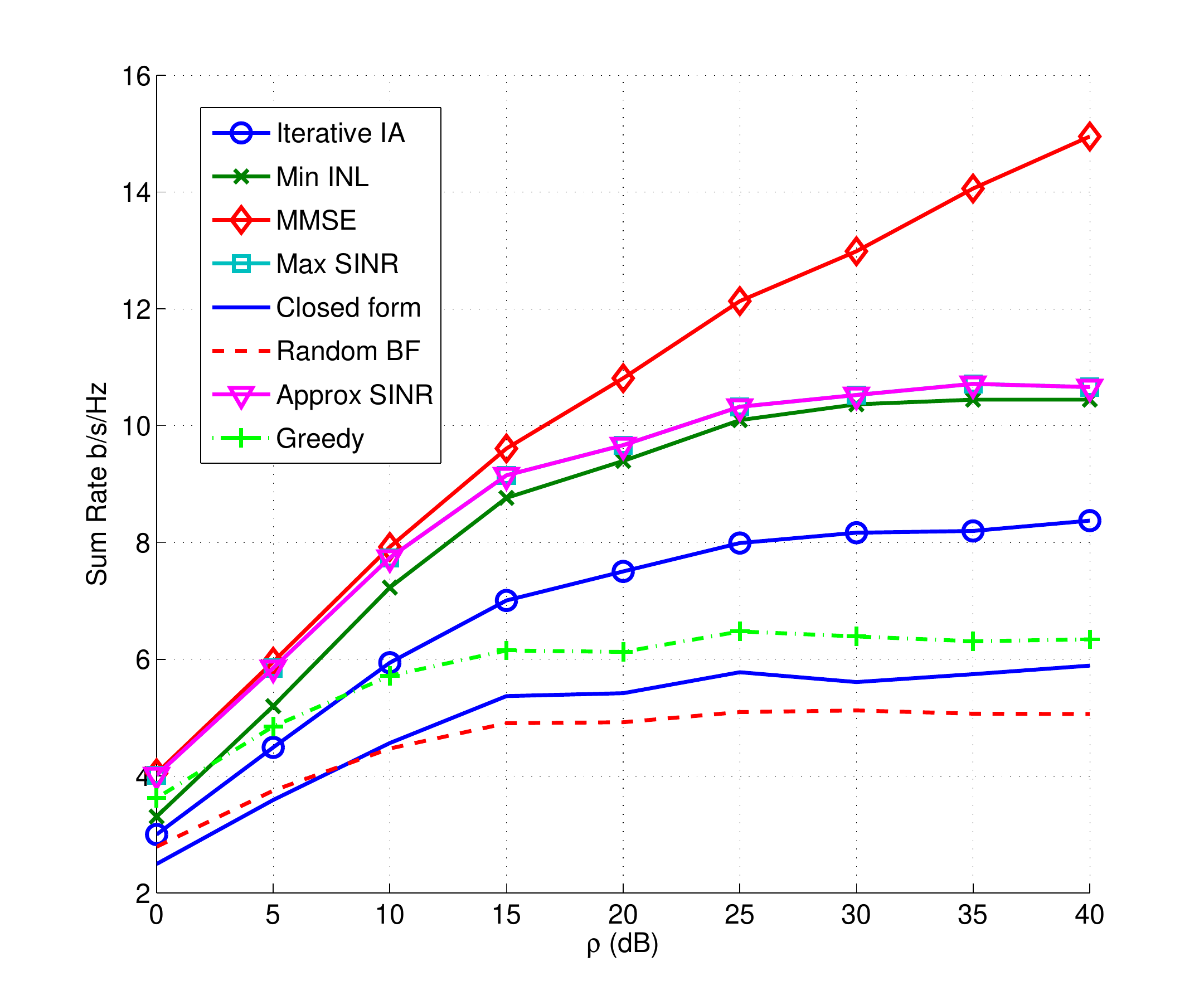}
\caption{Sum rate vs. $\rho_k=\rho_E=\rho$ for each algorithm discussed in Section~\ref{sec:algorithms} for the case where a rank-one uncoordinated interferer
is introduced into the (2,2,3) network with $S=1$ stream per user. The interferer's transmit power is scaled with the transmitters in the
network so the degrees of freedom are reduced, and the network is interference-limited at high values of $\rho$. The MMSE algorithm is an exception because it
has a power inequality constraint on its precoders and can thus allow two transmitters to turn off, giving the remaining transmitter one degree of freedom, so the
sum capacity scales linearly with $\rho$.}
\label{fig:if_scaled}
\end{figure}

Figure~\ref{fig:if_scaled} illustrates results for the rank-one interferer scenario for each algorithm 
discussed in Section~\ref{sec:algorithms} with $K=3$ users, $M=N=2$ antennas at each node, and $S=1$ stream being transmitted 
between each transmit/receive pair for $\rho_k=\rho_E=\rho$, $\forall k$ and $\alpha=1$. That is, the transmit power is equal at all transmitters, including
the uncoordinated interferer, and the path loss and fading statistics are identical on all links.

The MMSE algorithm has higher degrees of freedom in this case because of its power inequality constraint on the precoders. This allows two transmitters to effectively shut
off while the third has a degree of freedom and can cancel the external interferer with its extra receive antenna. This shows the flexibility of the MMSE design.
Other than MMSE, the max-SINR algorithm outperforms the others in the power ranges considered. Note that, although on average the max SINR 
algorithm and the approximate max SINR algorithm
have nearly identical performance, for any given channel realization they may have very different sum rates. IA performs the worst of all four iterative algorithms
since it is neglecting the uncoordinated interference. At high $\rho$, considering the colored noise in the algorithm objective results in a roughly 20\% increase
in sum rate for this scenario. Note that the two best-performing algorithms, MMSE and max-SINR, do not have orthogonal precoders and thus may be
more complex to implement in a real system with feedback requirements. With its orthogonal design and improved performance over IA,
the min-INL algorithm is a good tradeoff between complexity and performance in this scenario.

\begin{figure}
\centering
\includegraphics[width=3.5in]{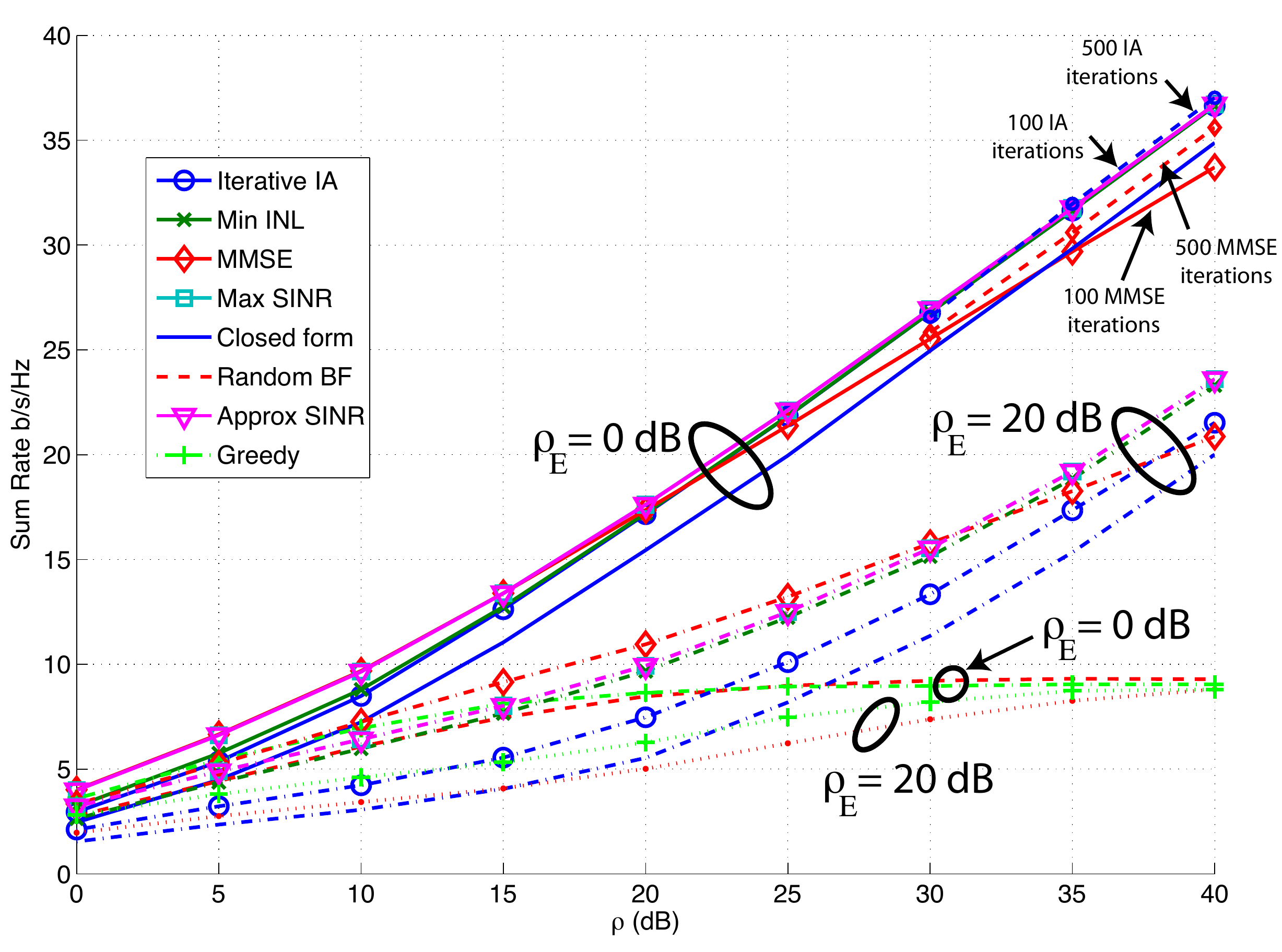}
\caption{Sum rate vs. $\rho_k=\rho$ for each algorithm discussed in Section~\ref{sec:algorithms} for the case where a rank-one uncoordinated 
interferer with fixed transmit power of $\rho_E = 0$ dB and $\rho_E = 20$ dB is introduced into the (2,2,3) network with $S=1$ stream per user. 
The degrees of freedom in this network are the same as if the interferer did not exist, and each algorithm, with the exception of random beamforming, performs
very similarly and exploits all the degrees of freedom in the network.}
\label{fig:if_fixed}
\end{figure}

Next, we keep the same scenario but with fixed uncoordinated interference power, so that the degrees of freedom are not reduced. 
Figure~\ref{fig:if_fixed} gives the results of this experiment. It shows that the uncoordinated interference, which is fixed at $\rho_E=0$ dB, has little
effect on the system, even at low $\rho_k=\rho$. The algorithms, except random beamforming, all scale at the same rate, and thus all exploit 
the maximum degrees of freedom in the network. For a fixed number of iterations, however, the MMSE algorithm does not scale, as it appears to require
more iterations to converge than the others at high $\rho$. In particular, as shown in Figure~\ref{fig:if_fixed}, when the MMSE design is run with 500 iterations, 
its performance approaches that of the rest of the designs, while the other algorithms benefit very little from the increase in iterations.
This is consistently seen in the rest of the simulations in this section. Analysis of this
longer convergence is left to future work. Finally, we note that iterative IA outperforms the closed-form solution because multiple IA solutions exist,
and iterative IA is better able to find the best one because of the multiple random initializations. If the closed-form algorithm is modified to explore 
multiple possible solutions, it would perform equally well in this case.

\begin{figure}
\centering
\includegraphics[width=3.5in]{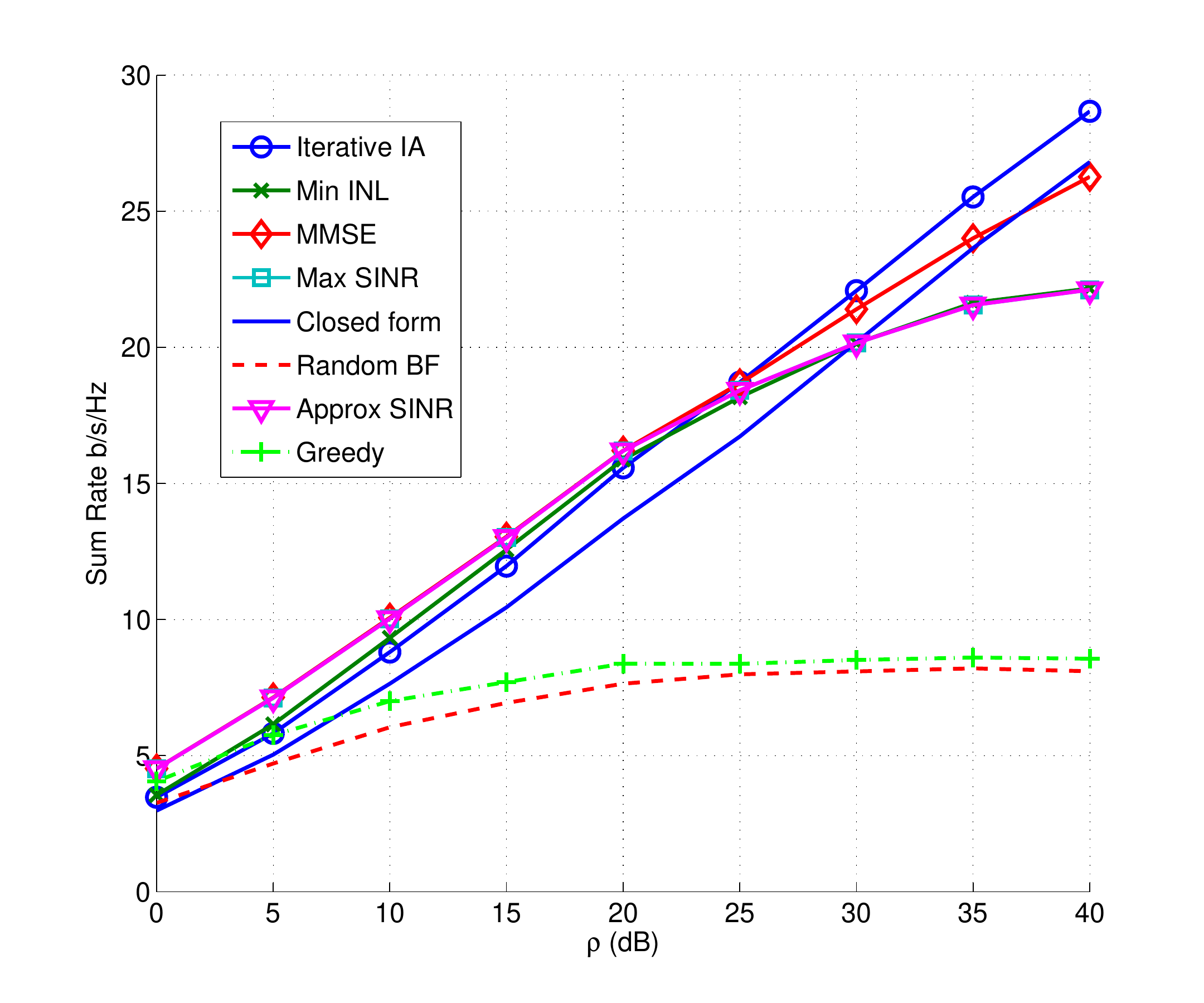}
\caption{Sum rate vs. $\rho_k=\rho_E=\rho$, with $\alpha_{k,E}=0$ for $k>1$ and $\alpha_{1,E}=1$, for each algorithm discussed in 
Section~\ref{sec:algorithms}. In this case, a rank-one uncoordinated  
interferer is sensed at only receiver 1 in the (2,2,3) MIMO interference channel with $S=1$. The interferer's transmit power is
scaled with the transmitters in the network so the degrees of freedom are reduced. The network is not interference-limited, however,
since only one receiver sees the interference.}
\label{fig:if_scaled_one}
\end{figure}

Now we remove the uncoordinated interferer from all but one receiver in the network, and allow that uncoordinated interference
power to scale with internal network transmit power, so that $\rho_k=\rho_E=\rho$, but $\alpha_{k,E}=0$ for $k>1$ and $\alpha_{1,E}=1$. 
Figure~\ref{fig:if_scaled_one} shows the results.
The maximum SINR and minimum INL algorithms suffer at high $\rho$ relative to pure interference
alignment and MMSE. This is because these algorithms see the large interferer at receiver $k=1$ as something to be overcome; these algorithms
are effectively concerned with the average performance of the network. IA is equally concerned about the average performance
but only in terms of coordinated interference, whereas MMSE has flexibility to overcome the interference by reducing transmit power of receiver 1. 
To maximize sum rate in this case, it appears one should either ignore the external interference or include a power inequality constraint in the precoder design.

\begin{figure}
\centering
\includegraphics[width=3.5in]{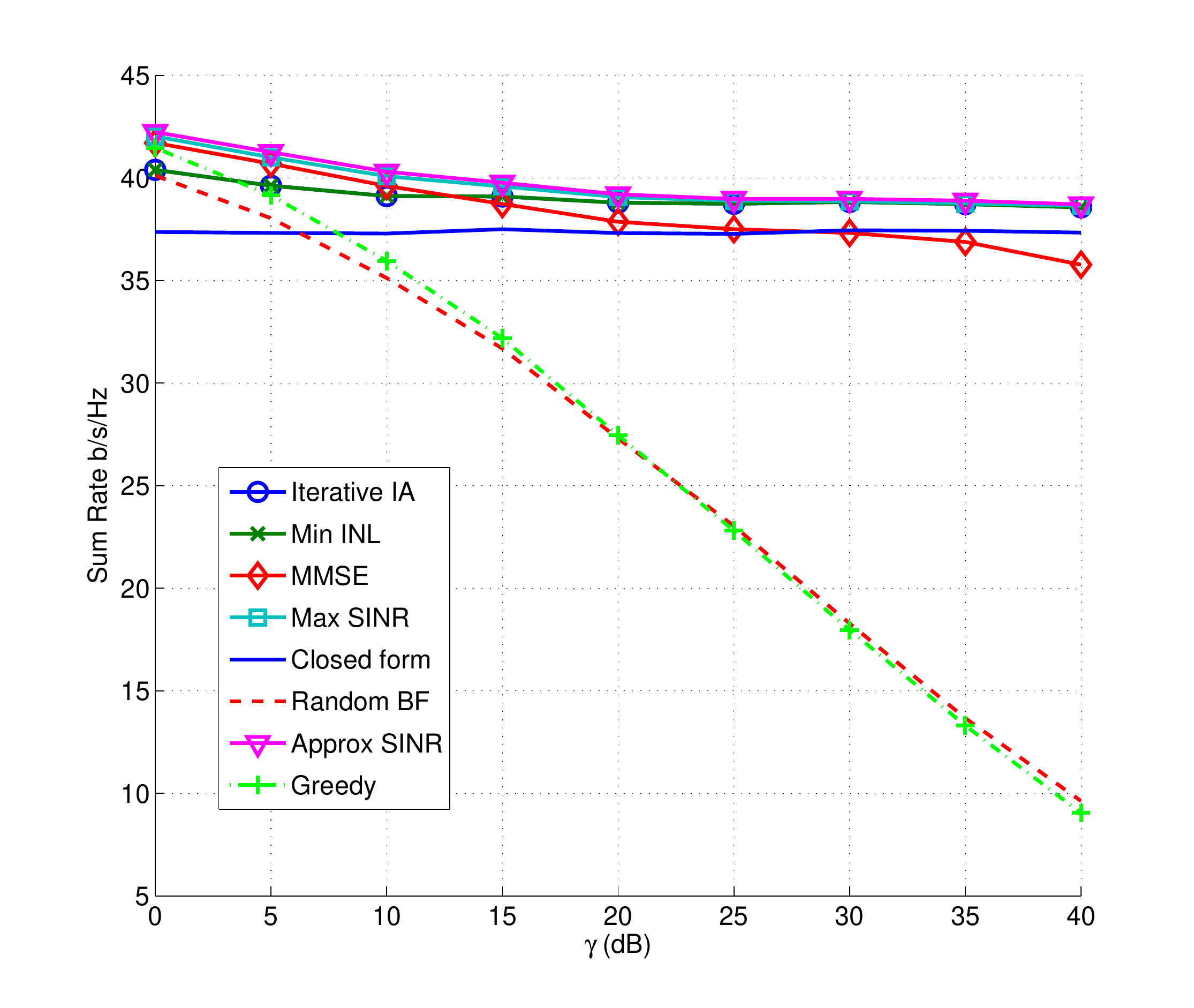}
\caption{Sum rate vs. $\alpha=\alpha_{k,\ell}$, $k\ne\ell$, for the $(2,2,3)$ MIMO interference channel with $S=1$ stream per user.
The iterative IA algorithm is not able to exploit reduced interference power, while all the other iterative algorithms can substantially
improve throughput. The SNR on the data links is fixed at $\gamma_{k,k} = 40$ dB.}
\label{fig:scale_if}
\end{figure}

We now turn to the case of no uncoordinated interference, considering only the conventional interference channel in isolation. In the first
experiment, the transmit power is kept fixed but the path loss coefficient $\alpha_{k,\ell}$ is varied on the interfering links ($k\ne\ell$) only. 
Figure~\ref{fig:scale_if} illustrates the results. The 
IA algorithm has constant throughput regardless of the interference path loss coefficient, but the other iterative algorithms
are able to exploit the decrease in interference, converging to IA when the interference power is high.

\section{Conclusions and Future Work}\label{sec:conclusion}
This paper has discussed the application and performance of iterative algorithms in the MIMO $K$-user constant-coefficient interference channel 
under various operating regimes. The convergence and optimality of the algorithms has been discussed, and similarities
between all of them have been derived. If an iterative solution for the interference channel is ever practical in a real system, it is unlikely that a direct 
interference alignment approach is desirable because of its suboptimality in environments where one or more links have little 
energy relative to the others. Instead, the max SINR or MMSE metrics are desirable in most environments because they 
flexibly adapt the solution between interference alignment (high interference power)
and SVD precoding (no interference, fixed number of streams), and the MMSE solution in particular has a transmit power inequality constraint. 
These algorithms, however, have relatively high implementation complexity because
of their nonorthogonality and lack of closed-form solutions at each step in general cases. In particular, the MMSE algorithm requires
some optimization for meeting the power constraint, and the max-SINR algorithm requires more channel state knowledge at each iteration than the others. 
The min-INL algorithm is a good tradeoff between the three algorithms, since it has improved performance over
IA in scenarios where there is uncoordinated interference or colored noise, but still has relatively low implementation complexity because of
its simpler solutions and orthogonal precoders.
Future work will focus on analyzing and reducing the overhead associated with solutions such as the ones presented in this paper.
Although some studies have been carried out on the application of interference alignment to a cellular 
network~\cite{SuhTse:Interference-Alignment-for-Cellular:08,CaiRamPap:Multiuser-MIMO-downlink:08,TreGui:Cellular-interference-alignment:09},
overhead and feedback analyses need to be performed to find out if the achievable gains are worth the effort.

%These algorithms can be improved still, however, because none of them directly maximize a throughput performance metric. 
%Indeed, the simulations show that any of them can be optimal under various finite-SNR operating conditions.
%Further, waterfilling or other dynamic power allocation among streams for a fixed user has been removed from the optimization in 
%these protocols and is worth investigating for the future.

%Since exploiting the time or frequency selectivity of the wireless channel appears to give significant throughput benefits for $K>3$,
%a comprehensive study on the solutions for a finite number of symbol extensions is warranted. 
%In particular, recent studies about the limited capacity of linear precoding IA with constant coefficients~\cite{YetJafKay:Feasibility-Conditions-for-Interference:09} 
%bring significant questions about the overhead issues associated with it.

\appendices
\section{Derivation of Mean Squared Error Minimization}\label{app:mmse}
\begin{IEEEproof}
For completeness, we restate the optimization from~(\ref{eq:mmse_obj}),
\begin{eqnarray}
\mathrm{minimize} & \mathcal{J}_{\rm MSE}\left(\{\bF_\ell\},\{\bG_k\}\right)\nonumber\\
\mathrm{subject~to} & \|\bF_\ell\|_F^2\le \rho_\ell, \ell\in\{1,\dots,K\}.
%\min_{\{\bG_k\},\{\bF_\ell\},\|\bF_\ell\|_F^2\le 1, \forall \ell} \mathcal{J}_{\rm MSE},
\label{eq:app_mmse_obj}
\end{eqnarray}
where $\mathcal{J}_{\rm MSE}$ is defined in~(\ref{eq:j_mmse}).
We use the Karush-Kuhn-Tucker conditions to solve the optimization at each step with all but one variable fixed.
The Lagrangian of~(\ref{eq:mmse_obj}) is 
\begin{eqnarray}
\mathcal{L} = \sum_{k=1}^K 
    {\rm tr}\left(\bG_k^*\left(\sum_{\ell=1}^K\bH_{k,\ell}\bF_\ell\bF_\ell^*\bH_{k,\ell}^* + \bR_k\right)\bG_k\right) -\nonumber\\
2\mathbb{R}\left\{{\rm tr}\left(\bG_k^*\bH_{k,k}\bF_k\right)\right\}
    + \sum_{\ell=1}^K\mu_\ell\left({\rm tr}\left(\bF_\ell^*\bF_\ell\right)-\rho_\ell\right),
\end{eqnarray}
where $\mu_\ell$ is the Lagrangian multiplier for the power constraint for precoder $\ell$.
The KKT conditions are 
\begin{eqnarray}
\nabla\mathcal{L} & = & {\bf 0}\label{eq:kkt1}\\
\mu_\ell\left({\rm tr}\left(\bF_\ell^*\bF_\ell\right)-1\right) & = & 0, \forall\ell\label{eq:kkt2}\\
{\rm tr}\left(\bF_\ell^*\bF_\ell\right) & \le & \rho_\ell, \forall\ell\label{eq:kkt3}\\
\mu_\ell & \ge & 0, \forall\ell\label{eq:kkt4}.
\end{eqnarray}
For fixed $\{\bF_\ell\}$ and $\{\mu_\ell\}$, $\{\bG_k\}$ can be found by solving $\nabla_{\bG_k}\mathcal{L}=\nabla_{\bG_k}\mathcal{J}_{\rm MSE} = 
{\bf 0}$ for 
$k\in\{1,\dots,K\}$ and the KKT conditions will be automatically met since there are no constraints on $\{\bG_k\}$. This yields
\begin{equation}
\bG_k = \left(\sum_{\ell=1}^K\bH_{k,\ell}\bF_\ell\bF_\ell^*\bH_{k,\ell}^* + \bR_k\right)^{-1}\bH_{k,k}\bF_k,
\end{equation}
In solving for $\{\bF_\ell\}$, we must ensure all of~(\ref{eq:kkt1})--(\ref{eq:kkt4}) are satisfied. To satisfy 
$\nabla_{\bF_\ell}\mathcal{L} = {\bf 0}$, we must have
\begin{equation}
\bF_\ell = \left(\mu_\ell\bI + \sum_{k=1}^K\bH_{k,\ell}^*\bG_k\bG_k^*\bH_{k,\ell}\right)^{-1}\bH_{\ell,\ell}^*\bG_\ell.
\end{equation}
If $\mu_\ell=0$ satisfies~(\ref{eq:kkt3}), then all the KKT conditions are satisfied and the optimal $\bF_\ell$ has been found for this step of the alternating
minimization. Otherwise, we must solve for $\mu_\ell>0$ such that $\left\|\bF_\ell\right\|_F = \rho_\ell$ to satisfy the KKT conditions. Although there is no
known closed-form solution for $\mu_\ell$ in this case~\cite{UluYen:Iterative-transmitter-and-receiver:04}, $\left\|\bF_\ell\right\|_F$ is a monotonically
decreasing function of $\mu_\ell$ for $\mu_\ell>0$, so simple one-dimensional searches such as the bisection method can be done to solve for $\{\mu_\ell\}$. 
% XXX TODO: equality constraints?
\end{IEEEproof}

%%%%%%%%%%%%%%%%%%%%%%%%%%%%%%%%%%%%%%%%%%%%%%%%%%%%%%%%%%%%%%%%%%%%%%%%%%%%%%%%%%%
\section{Derivation of SINR Maximization}\label{app:sinr}
\begin{IEEEproof}
For completeness, we restate the optimization from~(\ref{eq:sinr_obj}),
\begin{eqnarray}
\mathrm{maximize} & \mathcal{J}_{\rm SINR}\left(\{{\bf f}_\ell^{(n)}\},\{\bg_k^{(n)}\}\right)\nonumber\\ 
\mathrm{subject~to} & \|{\bf f}_\ell^{(n)}\|^2 = \frac{\rho_\ell}{S_\ell}, \forall n,\ell.
%\min_{\substack{\{\bg_k^{(n)}\},\{{\bf f}_\ell^{(n)}\}\\ 
%    \|{\bf f}_\ell^{(n)}\|^2 = 1/S_\ell, \forall n\in\{1,\dots,S_\ell\}, \ell\in\{1,\dots,K\}}} \mathcal{J}_{\rm SINR},
\label{eq:app_sinr_obj}
\end{eqnarray}
where $\mathcal{J}_{\rm SINR}$ is defined in~(\ref{eq:j_sinr}).
The optimization is performed on the columns of each precoder $\bF_\ell$ and spatial equalizer $\bG_k$.
In solving for the $n$th column of precoder $\bF_\ell$, we hold fixed every other precoder $\bF_k$, $k\ne\ell$, and 
other columns ${\bf f}_\ell^{(w)}$, $w\ne n$ of precoder $\bF_\ell$, as well as every receive combining matrix $\bG_k$, $\forall k$.
The objective of~(\ref{eq:j_sinr}) can be rewritten as
\begin{equation}
\mathcal{J}_{\rm SINR} = \frac{\sum_{\ell=1}^K\sum_{n=1}^{S_k}{\bf f}_\ell^{(n)*}\left(\bC_\ell^{(n)}+
  S_\ell r_\ell^{(n)}\bI\right){\bf f}_\ell^{(n)}}
%  \sum_{\substack{w=1\\ w\ne n}}^{S_\ell}S_\ell\left|\bg_\ell^{(w)*}\bH_{\ell,\ell}{\bf f}_\ell^{(w)}\right|^2\bI + 
%  \sum_{\substack{k=1\\ k\ne\ell}}^K\sum_{m=1}^{S_k}S_\ell\left|\bg_k^{(m)*}\bH_{k,k}{\bf f}_k^{(m)}\right|^2\bI\right)
 {\sum_{\ell=1}^K\sum_{n=1}^{S_k}{\bf f}_\ell^{(n)*}\left(S_\ell q_\ell^{(n)}\bI + 
 \bA_\ell^{(n)}+\bB_\ell^{(n)}\right){\bf f}_\ell^{(n)}},
\label{eq:app_sinr_expanded}
\end{equation}
where
\begin{eqnarray}
\bA_\ell^{(n)} & = & \sum_{\substack{w=1\\ w\ne n}}^{S_\ell}\bH_{\ell,\ell}^*\bg_\ell^{(w)}\bg_\ell^{(w)*}\bH_{\ell,\ell}\\
\bB_\ell^{(n)} & = & \sum_{\substack{k=1\\ k\ne\ell}}^K\sum_{m=1}^{S_k}\bH_{k,\ell}^*\bg_k^{(m)}\bg_k^{(m)*}\bH_{k,\ell}\\
\bC_\ell^{(n)} & = & \bH_{\ell,\ell}^*\bg_\ell^{(n)}\bg_\ell^{(n)*}\bH_{\ell,\ell}
\end{eqnarray}
and $q_\ell^{(n)}$ is the remaining summation terms in the denominator of~(\ref{eq:j_sinr}) that are independent of ${\bf f}_\ell^{(n)}$,
contracted here for brevity. Similarly, $r_\ell^{(n)}$ is the remaining summation terms in the numerator of~(\ref{eq:j_sinr}) that are
independent of ${\bf f}_\ell^{(n)}$. The function in~(\ref{eq:app_sinr_expanded}) is the generalized Rayleigh quotient which is well 
known to be solved by the generalized eigen-vector of the numerator and denominator matrices,
\begin{eqnarray}
{\bf f}_\ell^{(n)} = \sqrt{\frac{\rho_\ell}{S_\ell}}\nu_{\rm max}\biggl(\biggl(~S_\ell q_\ell^{(n)}\bI + 
  \bA_\ell^{(n)} + \bB_\ell^{(n)}\biggr)^{-1}\nonumber\\
  \biggl(\bC_\ell^{(n)} + S_\ell r_\ell^{(n)}\bI\biggr) \biggr).
%{\bf f}_\ell^{(n)} = \frac{1}{\sqrt{S_\ell}}\tilde{\nu}_{\rm max}\left(\bH_{\ell,\ell}^*\bg_\ell^{(n)}\bg_\ell^{(n)*}\bH_{\ell,\ell} + 
%  S_\ell r_\ell^{(n)}\bI,~S_\ell q_\ell^{(n)}\bI + 
%  \sum_{\substack{w=1\\ w\ne n}}^{S_\ell}\bH_{\ell,\ell}^*\bg_\ell^{(w)}\bg_\ell^{(w)*}\bH_{\ell,\ell} + 
%  \sum_{\substack{k=1\\ k\ne\ell}}^K\sum_{m=1}^{S_k}\bH_{k,\ell}^*\bg_k^{(m)}\bg_k^{(m)*}\bH_{k,\ell}\right).
\end{eqnarray}
The derivation of the receive combiner columns follows the same structure as the precoders, but with a unit-norm constraint on the columns
for simplicity (this removes the $S_\ell$ multipliers in front of $q_\ell^{(n)}$ and $r_\ell^{(n)}$). 
Here we define $\hat{r}_k^{(n)}$ to be the terms in the numerator of~(\ref{eq:j_sinr}) independent of $\bg_k^{(n)}$ and similarly
for $\hat{q}_k^{(n)}$ in the denominator. The solution for the $n$th column of the $k$th receive combiner is then
\begin{eqnarray}
\bg_k^{(n)} = \nu_{\rm max}\Biggl(\Biggl(~\hat{q}_k^{(n)}\bI + 
  \sum_{\substack{w=1\\ w\ne n}}^{S_k}\bH_{k,k}{\bf f}_k^{(w)}{\bf f}_k^{(w)*}\bH_{k,k}^* +\nonumber\\ 
  \sum_{\substack{\ell=1\\ \ell\ne k}}^K\sum_{m=1}^{S_\ell}\bH_{k,\ell}{\bf f}_\ell^{(m)}{\bf f}_\ell^{(m)*}\bH_{k,\ell}^*\Biggr)^{-1}\nonumber\\
  \left(\bH_{k,k}{\bf f}_k^{(n)}{\bf f}_k^{(n)*}\bH_{k,k}^* + 
  \hat{r}_k^{(n)}\bI\right)\Biggr).
%\bg_k^{(n)} = \tilde{\nu}_{\rm max}\left(\bH_{k,k}{\bf f}_k^{(n)}{\bf f}_k^{(n)*}\bH_{k,k}^* + 
%  \hat{r}_k^{(n)}\bI,~\hat{q}_k^{(n)}\bI + 
%  \sum_{\substack{w=1\\ w\ne n}}^{S_k}\bH_{k,k}{\bf f}_k^{(w)}{\bf f}_k^{(w)*}\bH_{\ell,\ell}^* + 
%  \sum_{\substack{\ell=1\\ \ell\ne k}}^K\sum_{m=1}^{S_\ell}\bH_{k,\ell}{\bf f}_\ell^{(m)}{\bf f}_\ell^{(m)*}\bH_{k,\ell}^*\right).
\end{eqnarray}
\end{IEEEproof}

\bibliographystyle{IEEEtran}
\bibliography{IEEEabrv,../../bibliographies/relay_coordination/relay_coordination,../huawei_consult/trunk/Multicell,../../bibliographies/mimo_relay/mimo_relay,../../bibliographies/mimo_basics/mimo_basics,../../bibliographies/interference_channel/interference_channel,../../bibliographies/information_theory/information_theory,../../bibliographies/ia_convergence/ia_convergence}

\newpage

\begin{IEEEbiography}[{\includegraphics[width=1in,height=1.25in,clip,keepaspectratio]{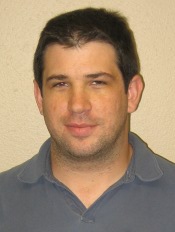}}]{Steven W. Peters}
 is a Ph.D. student at the University of Texas at Austin and Co-Founder and CEO of Kuma Signals, LLC. 
He received B.S.~degrees in electrical engineering and computer engineering
from the Illinois Institute of Technology in 2005 and an M.S.E.~degree in electrical engineering from the University
of Texas at Austin in 2007. From 2005--2007 he was a research assistant at the Applied Research Laboratories, where he worked
on physical layer and coding design for ground-wave high frequency transhorizon communication systems. He has served as a consultant 
to several companies working on wireless system design and standards. 
His research interests include interference mitigation techniques in wireless networks, cooperative communication, and MIMO.
\end{IEEEbiography}
\begin{IEEEbiography}[{\includegraphics[width=1in,height=1.25in,clip,keepaspectratio]{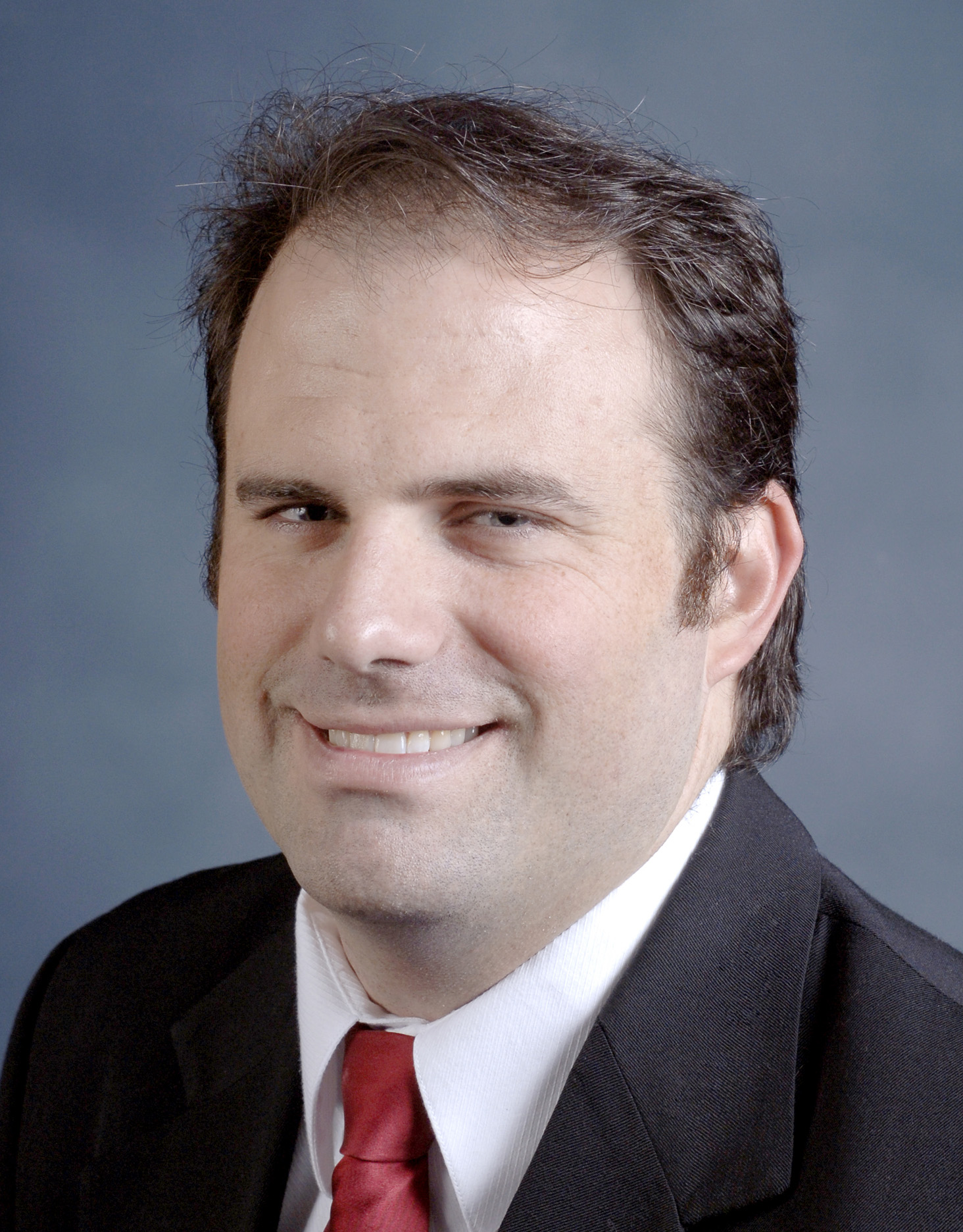}}]{Robert W. Heath, Jr.}
 (S'96 - M'01 - SM'06)  received the B.S. and M.S. degrees
from the University of Virginia, Charlottesville, VA, in 1996 and 1997 respectively, and the Ph.D. from Stanford University, Stanford, CA, in 2002, all in electrical engineering.

From 1998 to 2001, he was a Senior Member of the Technical Staff then a Senior Consultant at Iospan Wireless Inc, San Jose, CA where he worked on the design and implementation of the physical and link layers of the first commercial MIMO-OFDM communication system. In 2003 he founded MIMO Wireless Inc, a consulting company dedicated to the advancement of MIMO technology. Since January 2002, he has been with the Department of Electrical and Computer Engineering at The University of Texas at Austin where he is currently an Associate Professor and Associate Director of the Wireless Networking and Communications Group. His research interests include several aspects of MIMO communication: limited feedback techniques, multihop networking,  multiuser MIMO, and multicell MIMO, as well as 60GHz communication techniques and multi-media signal processing.

Dr. Heath has been an Editor for the IEEE Transactions on Communication and an Associate Editor for the IEEE Transactions on Vehicular Technology. He is a member of the Signal Processing for Communications Technical Committee in the IEEE Signal Processing Society and is the Vice Chair of the IEEE COMSOC Communications Technical Theory Committee. He was a technical co-chair for the 2007 Fall Vehicular Technology Conference, general chair of the 2008 Communication Theory Workshop, general co-chair, technical co-chair and co-organizer of the 2009 IEEE Signal Processing for Wireless Communications Workshop, local co-organizer for the 2009 IEEE CAMSAP Conference, and was technical co-chair for the 2010 IEEE International Symposium on Information Theory. He is the recipient of the David and Doris Lybarger Endowed Faculty Fellowship in Engineering. He is a licensed Amateur Radio Operator and is a registered Professional Engineer in Texas.
\end{IEEEbiography}

\end{document}